\documentclass[pdflatex,sn-mathphys]{sn-jnl}% Math and Physical Sciences Reference Style
\pdfoutput=1
\usepackage[utf8]{inputenc}
\usepackage{amsmath,amsfonts,amssymb,mathtools,amsthm}
\usepackage{graphicx,float}
\usepackage{caption}
\usepackage{subcaption}
\usepackage{xcolor}
\usepackage{listings}
\usepackage{adjustbox}
\usepackage{tablefootnote}
\usepackage{multicol,multirow}
\jyear{2021}%

\theoremstyle{thmstyleone}%
\newtheorem{theorem}{Theorem}%  meant for continuous numbers
\newtheorem{proposition}[theorem]{Proposition}% 

\theoremstyle{thmstyletwo}%

\theoremstyle{thmstylethree}%

\makeatletter
\newcommand\notsotiny{\@setfontsize\notsotiny\@vipt\@viipt}
\makeatother

\makeatletter
\newcommand\footnoteref[1]{\protected@xdef\@thefnmark{\ref{#1}}\@footnotemark}
\makeatother

\begin{document}

\title{New designs for Bayesian adaptive cluster-randomized trials}


\author*[1]{\fnm{Junwei} \sur{Shen}}%\email{junwei.shen@mail.mcgill.ca}

\author[1]{\fnm{Shirin} \sur{Golchi}}%\email{junwei.shen@mail.mcgill.ca}

\author[1]{\fnm{Erica EM} \sur{Moodie}}%\

\author[2,3]{\fnm{David} \sur{Benrimoh}}%\


\affil*[1]{\orgdiv{Department of Epidemiology, Biostatistics and Occupational Health}, \orgname{McGill University}%, \orgaddress{\street{1020 Pine Avenue West}, \city{Montreal}, \postcode{H3A 1A2}, \state{QC}, \country{Canada}}
}

\affil[2]{%\orgdiv{Department of Epidemiology, Biostatistics and Occupational Health}, 
\orgname{Aifred Health, \orgaddress{\city{Montreal}, \country{Canada}}}}
%, \orgaddress{\street{1020 Pine Avenue West}, \city{Montreal}, \postcode{H3A 1A2}, \state{QC}, \country{Canada}}}

%\affil[2]{\orgdiv{Department}, \orgname{Organization}, \orgaddress{\street{Street}, \city{City}, \postcode{10587}, \state{State}, \country{Country}}}

\affil[3]{\orgdiv{Department of Psychiatry}, \orgname{McGill University}%, \orgaddress{\street{Street}, \city{City}, \postcode{610101}, \state{State}, \country{Country}}
}

%\affil[3]{\orgdiv{Department}, \orgname{Organization}, \orgaddress{\street{Street}, \city{City}, \postcode{610101}, \state{State}, \country{Country}}}


\abstract{Adaptive approaches, allowing for more flexible trial design, have been proposed for individually randomized trials to save time or reduce sample size. However, adaptive designs for cluster-randomized trials in which groups of participants rather than individuals are randomized to treatment arms are less common. Motivated by a cluster-randomized trial designed to assess the effectiveness of a machine-learning based clinical decision support system for physicians treating patients with depression, two Bayesian adaptive designs for cluster-randomized trials are proposed to allow for early stopping for efficacy at pre-planned interim analyses. The difference between the two designs lies in the way that participants are sequentially recruited. Given a maximum number of clusters as well as maximum cluster size allowed in the trial, one design sequentially recruits clusters with the given maximum cluster size, while the other recruits all clusters at the beginning of the trial but sequentially enrolls individual participants until the trial is stopped early for efficacy or the final analysis has been reached. The design operating characteristics are explored via simulations for a variety of scenarios and two outcome types for the two designs. The simulation results show that for different outcomes the design choice may be different. We make recommendations for designs of Bayesian adaptive cluster-randomized trial based on the simulation results. }




\maketitle

\section{Introduction}\label{sec1}
Randomized controlled trials, which can ensure that subjects assigned to each treatment group are comparable with respect to all characteristics of interest to draw a causal conclusion, have played an essential role in evaluating the effectiveness of interventions \cite{rct}. In most trials designed to assess the effect of a drug or a treatment, individual participants are randomly allocated to each treatment arm. However, some interventions naturally operate at a group level, or target either a social network or physical environment. For example, in a trial assessing the clinical utility, safety, and potential effectiveness of a machine-learning based clinical decision support system (CDSS) developed by Aifred Health \cite{aifred1,aifred4, aifred5, aifred6}, the decision support tool is designed for physicians, naturally forming clusters of individual patients being treated by that physician \cite{aifred2,aifred3}. For these types of interventions, a population-level effect is of more interests to researchers. 

In addition, randomizing by individuals in this setting may lead to a contamination effect between trial arms, as physicians might struggle to treat patients from different treatment arms strictly differently. Contamination can cause dilution bias and affect the reliability and validity of the study. One way to reduce the possibility of contamination is to randomize by physicians which act as `clusters' \cite{crtintro3, crtintro5}. Thus, it may not be advisable to randomize individual participants to different treatment arms, and groups of subjects being treated by their clinicians can instead be randomly assigned to the treatment arms in what is known as a \textit{cluster-randomized trial} \cite{crt,crtintro1,crtintro2,crtintro5}. 

%In this setting, it is not feasible to randomize individual participants to different treatment arms, and groups of subjects will be instead randomly assigned to the treatment arms in what is known as \textit{cluster-randomized trial} \cite{crt,crtintro1,crtintro2,crtintro5}. 

%Another motivating factor for cluster-randomized trials is to avoid contamination between trial arms \cite{crt, crtintro1, crtintro3}. 
%In our example trial, since the CDA is used by physicians, physicians as the cluster are randomized to the intervention or control arm of the trial. %The use of CDA is described in section \ref{sec2}.
%If individual patients are instead randomized, the physicians might struggle to treat patients from different groups strictly differently. 
%Contamination can cause dilution bias and affect the reliability as well as validity of the study. Compared with an individually randomized trial,  cluster-randomized trials reduce the possibility of contamination, and therefore avoid the dilution bias \cite{crtintro3, crtintro5}.

Due to the difference in randomization unit, the design and analysis of cluster-randomized trials differ from individually randomized trials \cite{crtintro6,crtintro7,crtintro8}. The independence assumption between participants' outcomes is  violated in a cluster-randomized trial since subjects from the same cluster tend to have more similar responses than subjects from a different cluster. Correlation between subjects within the same cluster, as measured by intra-cluster correlation coefficient (ICC), should be considered carefully when analyzing data from cluster randomized trials.  The correlation reduces the variability of responses in a clustered sample and thus reduces the statistical power to detect true differences between treatment arms relative to trials that randomize the same number of individuals \cite{icc,icc2}. 

A natural way to improve trial efficiency is to incorporate adaptive features into the design which allow for planned adjustments to the trial design  %and/or statistical procedures of the trial 
after its initiation without undermining the validity and integrity of the trial \cite{adaptive1}. %Trials can be altered in midcourse in adaptive designs based on accrued information, and therefore are more flexible than fixed designs.  
In addition to flexibility and efficiency, adaptive designs are attractive to clinical scientists because they may reflect medical practice in the real world and they are ethical with respect to the need to determine and monitor efficacy as well as safety of the treatment \cite{adaptive2}.  Commonly used adaptive designs in clinical trials include, but are not limited to, adaptive randomization, group sequential design, and stopping rules \cite{adaptive2,adaptive3}.
%, but are not limited to: (a) an adaptive randomization design, (b) a group sequential design, (c) a sample size re-estimation design, (d) a drop-the-loser design, (e) an adaptive dose finding (e.g., dose escalation) design, (f) a biomarker-adaptive design, (g) an adaptive treatment-switching design, (h) a hypothesis-adaptive design, (l) an adaptive seamless phase II/III trial design, and (m) a multiple adaptive design. The details of these adaptive designs are described in \cite{adaptive2,adaptive3}. 

Adaptive designs naturally fit into the Bayesian framework as %due to the feature %of adaptive clinical trials 
that results or estimates are continually updated based on the accumulated information from interim data %has made the Bayesian framework especially appealing 
%adaptive designs naturally fit Bayesian paradigm
\cite{bat1}. %In the Bayesian paradigm, unknown parameters are considered as random quantities from some prior distribution, and their inferences are drawn from the corresponding posterior distribution using all available data.  
%and obtained from Bayes' theorem. 
Adaptive designs based on Bayesian approaches have been extensively studied in recent years \cite{bat,bat2,bat3,bat4,bat5,bat6}. However, most adaptive designs, regardless of whether frequentist or Bayesian, focus on individually randomized trials.
Adaptive designs for cluster-randomized trials are less common and the incorporation of adaptive features poses significant statistical challenges. Some specific adaptive features such as sample size re-estimation and group sequential design have been proposed in combination with cluster-randomized trials \cite{battry1,battry2,battry3,battry4}. However, no formal statistical design of Bayesian adaptive cluster-randomized trials has been developed. We address this gap by proposing two Bayesian adaptive designs for cluster-randomized trials.

The organization this paper is as follows. The motivating trial is briefly described in section \ref{sec2}, followed by the development of two Bayesian adaptive designs and models for continuous and binary outcomes in section \ref{sec3}. Simulation studies are carried out in section \ref{sec4}, assessing the performance of the two proposed designs across a range of scenarios. The paper concludes in section \ref{sec5}.

\section{Motivating setting} \label{sec2}
Aifred Health \cite{aifred1, aifred4, aifred5, aifred6} % (\url{https://www.aifredhealth.com/}) 
has designed a machine-learning based clinical decision support system (CDSS) for physicians treating patients with depression. Patient characteristics of interests such as sociodemographic information, clinical information and medical history are input into the CDSS. The CDSS then, using a deep learning model, outputs the predicted efficacy for a number of possible treatments for that patient \cite{aifred2, aifred3}.  %a list of all possible treatments, with each treatment associated with a predicted efficacy. 
Treatments are ordered by efficacy and presented to the physician when they reach the treatment selection step of a clinical algorithm based on best practice guidelines \cite{aifred4, aifred7}. Physicians with the CDSS can, on an individual patient basis, decide whether or not to use the information presented by CDSS as part of their medical decision-making.

%Aifred Health has been validating this AI-driven treatment recommendation tool. One 
An important step is to establish the clinical utility, safety, and potential effectiveness of a tool such as the CDSS. Practically, an intervention such as the CDSS must be delivered at the physician level, such that a cluster-randomized design would be appealing for the reasons described above. %of interest. % Therefore, a  patient-and-rater blinded and partially physician blinded, physician randomized,
%prospective controlled trial has been designed. 
Participating physicians could be randomized, with %to either intervention or control group based on factors that could influence technology usage or patient population. Such factors could include the length of time in practice, specialty (psychiatry or family medicine/primary care) and practice setting (community or tertiary care). Patients 
patients recruited from physicians' usual practices in order to approximate the real-world clinical conditions and populations as closely as possible. %and they will recruit patients from their usual practices. 
%Physicians in intervention group  who diagnose a patient with depression will have the choice to use the CDA after baseline measurement. Physicians in the control group will proceed according to the usual guideline-informed practice with the benefit of patient symptom questionnaires provided to them by paper or secure, site-approved electronic means.

Typical outcomes in a study of depression in such a trial could be a continuous measure of symptoms (e.g., a visual analog scale, the Quick Inventory of Depressive Symptomatology \cite{depressionmeasure4}, the 9-question depression scale from the Patient Health Questionnaire \cite{depressionmeasure5}, the World Health Organization Disability Assessment Schedule 2.0 \cite{depressionmeasure6}, etc.) or a binary measure of treatment response, minimum clinically significant change in symptoms, or remission, perhaps defined by a dichotomization of a standard depression score \cite{depressionmeasure1,depressionmeasure2,depressionmeasure3}. The importance of mental health treatment and the often relatively slow rate of patient accrual motivate the use of a Bayesian adaptive trial design to ensure adequate sample size and the possibility of early termination due to treatment effectiveness.

\section{Methods}\label{sec3}

%In this section, two Bayesian adaptive cluster-randomized designs which incorporate stopping rules for efficacy at interim analysis  are proposed \textcolor{red}{and standard models for interim analysis are described to illustrate how the designs proceed.}

\subsection{Two Bayesian adaptive cluster-randomized designs} \label{subsec31}
Suppose that the maximum number of clusters and maximum cluster size are equal across the two treatment arms.  Let $K$, $n$, $m$ be the number of interim analyses (not including the final analysis), the maximum number of clusters for each treatment arm, and the maximum cluster size, respectively. For simplicity of exposition, we will assume that all clusters enroll the same number of participants. Two Bayesian designs, design 1 and design 2 in the remainder of this paper, are developed to sequentially enroll participants and analyze interim data in different ways. 
 
 %\subsubsection{Design 1} 
In design 1,  $[\frac{n}{K+1}]$ clusters enter the trial at the start where $[x]$ denotes the largest integer not exceeding $x$, and $m$ individual participants will be enrolled for each cluster. At the subsequent analysis point, if the accumulated information up to the current analysis is sufficient to conclude the efficacy of the treatment or the final analysis has been reached, the trial will be terminated. Otherwise, another $[\frac{n}{K+1}]$ new clusters will be enrolled, and $m$ individual participants will be recruited for each new cluster. The trial then proceeds to the next analysis with new samples. This procedure is repeated until termination.
 
 %\subsubsection{Design 2} 
In design 2, at the beginning of the trial, all $n$ clusters enter the trial, but only $[\frac{m}{K+1}]$ individual participants are enrolled for each cluster. At the subsequent analysis point, if the trial is not to be terminated, another $[\frac{m}{K+1}]$ individual participants are recruited for the same $n$ clusters. The trial then proceeds to the next analysis. This procedure is repeated until termination.

\begin{figure}[H]
    \centering
    \includegraphics[width=\textwidth]{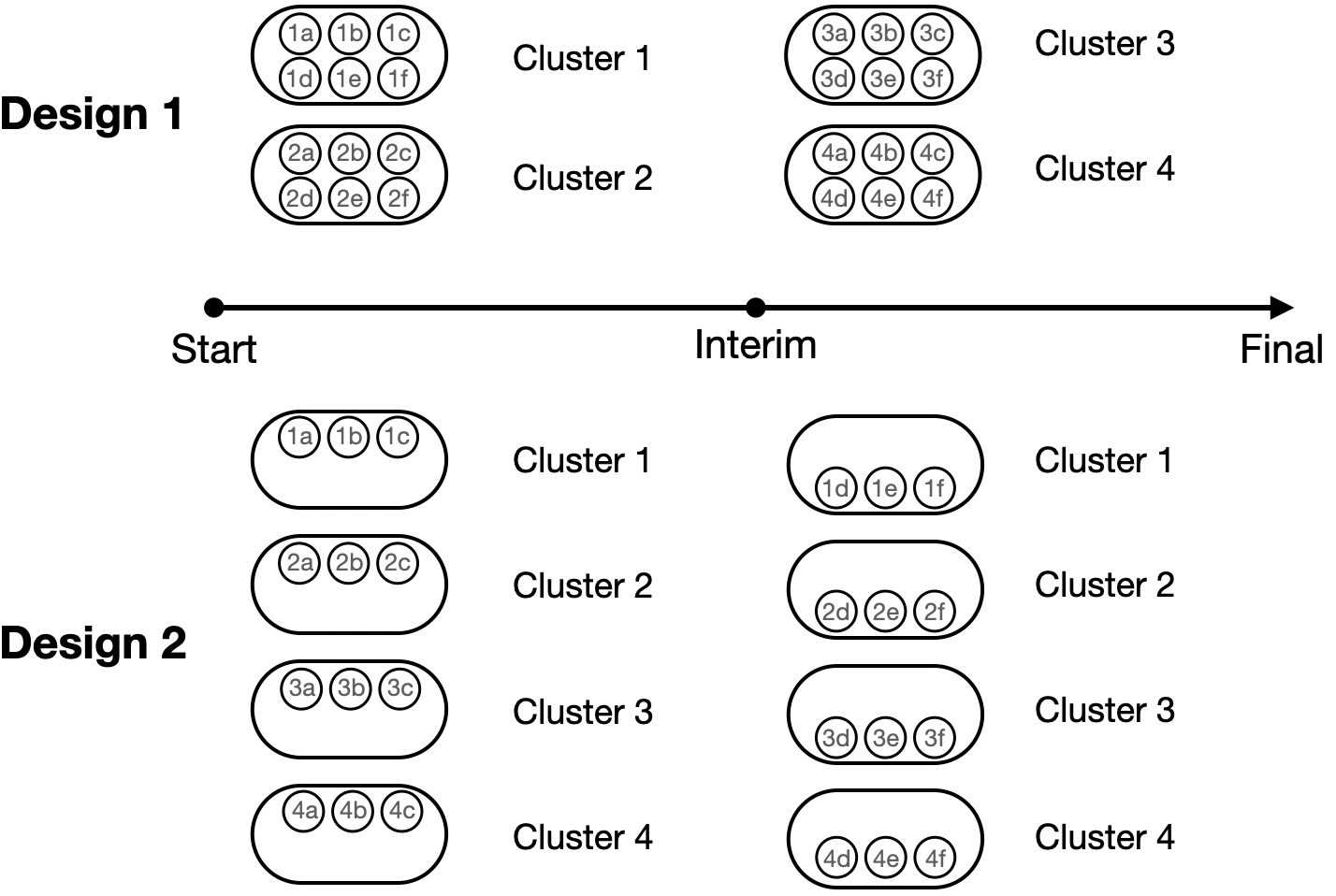}
    \caption{An illustration of the two designs for one treatment arm when %$K=1$, $n=4$ and $m=6$. 
    there is only one pre-planned interim analysis, and at most four clusters with a total of six individual participants within each cluster over the course of the trial.% maximum number of clusters
    The black labelled circles represent different individual participants for each cluster.}
    \label{fig:example}
\end{figure}

The fundamental difference between the two designs lies in the way that participants are sequentially recruited. In design 1, the \textit{clusters} are sequentially enrolled, and individual participants for each cluster are recruited all at once. In design 2, all clusters are enrolled at one time, but the \textit{individual participants} for each cluster are sequentially enrolled. For illustration, consider an example for one treatment arm with $K=1$, $n=4$ and $m=6$. That is, there is only one interim analysis planned (thus two analyses in total, including the final analysis). The maximum number of clusters is 4, and the maximum cluster size is 6. Figure \ref{fig:example} gives a graphical illustration. The black labelled circles represent different individual participants for the corresponding clusters. In design 1, clusters 1 and 2 will be first enrolled, and for each, six individual participants will be recruited. The interim analysis is based on data from clusters 1 and 2. If we decide to continue the trial, then we will further recruit clusters 3 and 4 and their respective six individual participants for the final analysis. The final analysis is based on data from all the four clusters. In design 2, all four clusters will be recruited at the start, but for each cluster, only three participants will be enrolled. If evidence based on the 12 individuals from 4 clusters is unable to conclude the efficacy of the intervention at the interim analysis, then an additional three individual participants for each cluster will be recruited, and the trial proceeds to the final analysis.

\subsection{Early stopping at interim analysis} \label{subsec32}
At each interim look, one should determine whether to stop the trial early or continue based on the interim result. The goal is to evaluate the efficacy of the treatment by testing the hypothesis
%\begin{equation}
%   H_0: \theta \leq \delta \quad \text{vs} \quad H_A: \theta > \delta
%\end{equation}
$$H_0: \theta \leq 0 \quad vs \quad H_A: \theta > 0,$$ 
where $\theta$ is the mean difference for continuous outcome and risk difference for binary outcome. %The efficacy of the treatment can be concluded if the posterior probability of $\theta$ exceeding minimal important difference is greater than a prespecified decision boundary. In other words, the trial can be stopped early for efficacy at the $k$-th interim analysis if 
The efficacy of the treatment can be concluded and the trial can be stopped early for efficacy at the $k$-th interim analysis if
\begin{align}
    P(\theta > \delta \vert D_k)>U,
    \label{eq:posterior}
\end{align}
%$$$$
where $\delta$ is the minimal important difference, $D_k$ is all the available data up to the $k$-th analysis and $U$ is the decision boundary.

\subsubsection{Continuous outcomes} \label{subsec321}
Based on the context of the motivating trial and without loss of generality, assume that a smaller value of the continuous outcome is preferred (as most depression and disability rating scales associate worse symptoms with larger total scores). Then, $\theta=\mu_c-\mu_t$ where $\mu_c$, $\mu_t$ are the mean outcome for control and treatment groups, respectively. %In the motivating trial, the continuous outcome can be some measures of symptoms or disability. 
We further assume that at the $k$-th analysis, $k=1, \ldots, K+1$, there are $n_k$ clusters with $m_k$ observations in each cluster. Let $Y_{ij}$  be the continuous outcome of the $i$-th subject in the $j$-th cluster at current analysis point, $j=1, \ldots, n_k$, $i=1, \ldots, m_k$. The normality assumption in cluster-randomized trials states 
\begin{align*}
    \mu_j \sim N(\mu, \sigma_B^2) \quad \text{and} \quad Y_{ij} \vert \mu_j \sim N(\mu_j, \sigma_W^2),
\end{align*}
where $\mu_j$ is the cluster-specific mean, $\mu$ is the population mean, $\sigma_W^2$ and $\sigma_B^2$ are within- and between-cluster variances respectively, and they can be related via the ICC, $\rho= \frac{\sigma_B^2}{\sigma_W^2+\sigma_B^2}$. It can be shown that the marginal distribution of $Y_{ij}$ is normal with mean $\mu$ and variance $\sigma_B^2+\sigma_W^2$. Let $Y=(Y_1^T,\ldots, Y_{n_k}^T)^T$ be the response vector where  $Y_j=(Y_{1j},\ldots, Y_{m_kj})^T$, for $j=1, \ldots, n_k$, and the covariance structure satisfies
\begin{align}
\label{eq1}
\begin{split}
    \text{Var}(Y_{ij})&=\sigma_B^2+\sigma_W^2\\
    \text{Cov}(Y_{ij}, Y_{sj})&=\sigma_B^2, \quad \forall i \neq s \\ 
    \text{Cov}(Y_{ij}, Y_{tl})&=0, \quad \forall j \neq l 
    \end{split}
\end{align}
so that $Y\sim \text{MVN}(\overrightarrow{\mu}, \Sigma)$ where $\overrightarrow{\mu}$ is a vector of $n_k \times m_k$ with all elements equal to $\mu$ and $\Sigma$ is a block matrix of the form 
$$\begin{pmatrix}
\Sigma_{Y_1} & \mathbf{0}  &\cdots & \mathbf{0}\\
\mathbf{0} & \Sigma_{Y_2}  & \cdots &\mathbf{0}\\
\vdots & \vdots & \ddots  &\mathbf{0}\\
\mathbf{0} & \mathbf{0} & \cdots &\Sigma_{Y_{n_k}}
\end{pmatrix}
$$
and $\Sigma_{Y_j}=\text{cov}(Y_j)$ specified in equation (\ref{eq1}), and MVN indicates a multivariate normal distribution.

In the Bayesian framework, assume a normal prior for $\mu$,
$$\mu \sim N(a,b^2)$$
where $a$ is the prior mean and $b^2$ is the prior variance. Then, using Bayes' theorem,  the posterior distribution for $\mu$ can be obtained:
$$\mu \vert Y \sim N\left(\frac{b^2\textbf{1}^T\Sigma^{-1}Y+a}{b^2\textbf{1}^T\Sigma^{-1}\textbf{1}+1},\frac{b^2}{b^2\textbf{1}^T\Sigma^{-1}\textbf{1}+1}\right).$$
The above result applies to both $\mu_t$ and $\mu_c$. Therefore the quantity of interest $P(\mu_c-\mu_t>\delta\vert Y)$ can be estimated by 
$$
\hat{\pi}=\frac{1}{M}\sum_{i=1}^MI\{\mu_{ci}^{post}-\mu_{ti}^{post}>\delta\}
$$
where $\mu_{ci}^{post}, \mu_{ti}^{post}$ are sampled from the corresponding posterior distributions of $\mu_c$ and $\mu_t$, and $M$ is the number of Monte Carlo samples drawn from the posterior. 

At any interim analysis, if $\hat{\pi}>U$, the trial is stopped early for efficacy. Otherwise, the trial continues to enroll clusters/participants and proceeds to the next analysis where the prior mean and variance for the next analysis are updated with the posterior mean and variance for the current analysis. These steps are repeated until either the trial is stopped early for efficacy or reaches the final analysis.

\subsubsection{Binary outcomes} \label{subsec322}
For binary outcomes, assume that a larger proportion is preferred (e.g., a larger proportion of patients meeting the criteria for treatment response). Let $\theta=\pi_t-\pi_c$ where $\pi_c$, $\pi_t$ are the population proportion for the control and treatment groups, respectively. %In the motivating trial, %the binary outcome could be, for example, the remission status, 
%several different binary outcomes could be employed as discussed in section \ref{sec2}. Then $\pi_c$ and $\pi_t$ could %can be remission rate be the rate of the binary outcome for the control and treatment group respectively.
For binary outcomes, the joint distribution of all observations cannot be obtained analytically and thus a tractable form of the posterior distribution of the parameters of interest is not available. Therefore, a hierarchical model is proposed as follows
\begin{align*}
    \pi_j &\sim \text{Beta}(\alpha,\beta)\\
       r_j \vert \pi_j &\sim \text{Binomial}(m_k, \pi_j)
\end{align*}
where $r_j$ is the number of events in the $j$-th cluster and $\pi_j$ is the cluster-specific proportion, $j=1, \ldots, n_k$. Then, under the model, the cluster-specific proportion $\pi_j$ have the mean-variance relationship  
\begin{align}
\label{eq2}
    \text{Var}(\pi_j)=E(\pi_j)(1-E(\pi_j))\frac{1}{\alpha+\beta+1}.
\end{align}
However, in cluster-randomized trials with binary outcomes, it is assumed that 
\begin{equation}
    \begin{split}
    E(\pi_j)&=\pi\\
    \label{eq3}
    \text{Var}(\pi_j)&=\rho \pi (1-\pi)
    \end{split}
\end{equation}
where $\pi$ is the population proportion and $\rho$ is the ICC.

To make (\ref{eq2}) and (\ref{eq3}) consistent, define two transformed parameters 
\begin{align*}
   % \rho&=\frac{1}{\alpha+\beta+1}\\
    \pi&=\frac{\alpha}{\alpha+\beta}\\
    v&=\alpha+\beta
\end{align*}
where  $\pi$ is exactly the mean of the Beta distribution and $v$ measures the information in the corresponding Beta distribution. Also, due to the consistency of (\ref{eq2}) and (\ref{eq3}), once $\rho$ is fixed or can be estimated, $v$ can also be determined through $v=\frac{1-\rho}{\rho}$. Thus, the only free parameter for the hierarchical model is $\pi$.

In the Bayesian framework, assume a uniform prior for $\pi$, 
$$ \pi \sim U[0,1].$$
Then the posterior probability $P(\pi_t-\pi_c>\delta \vert R)$ can be approximated by drawing posterior samples using Markov Chain Monte Carlo (MCMC) implemented in, say \texttt{RStan} \cite{stan,rstan}, where $R=(r_{1c}, \cdots, r_{n_k,c}, r_{1t}, \cdots, r_{n_k,t})$ includes the number of events per cluster within each treatment group.

\section{Simulation studies} \label{sec4}

The false positive rate and power cannot be obtained analytically in Bayesian adaptive trials since the sampling distributions of the test statistics (i.e., posterior probability statements (\ref{eq:posterior}) in section \ref{subsec32}) are not known. Therefore simulation studies are required to specify the decision boundaries and other design characteristics \cite{bat}. For both outcomes, a single interim analysis was explored first; two interim analyses were then  investigated. %For single interim analysis, we fixed the cluster size as 8 while for multiple interim analyses a larger cluster size as 16 was also explored. 
The minimal important difference was set as 0 in the simulation but it is straightforward to generalize to other values. For each scenario, 500 simulation replications were performed. The performance of designs was compared based on false positive rate and power. The false positive rate is estimated as
\begin{align*}
    \text{False positive rate} = \frac{\text{Number of times the null hypothesis is falsely rejected }}{\text{Number of simulation runs}},
\end{align*}
where falsely rejecting the null hypothesis means that for some $k\leq K+1$, $$P_{\theta=0}(\hat{\theta}_k>\delta \vert D_k)>U$$
for $\hat{\theta}_k$ the estimated mean difference for continuous outcome or estimated risk difference for binary outcome at the $k$-th analysis, $\theta$ is the true mean or risk difference, $\delta$ is minimal important difference, and $\delta=0$ in our simulation. $D_k$ is all the available data up to the $k$-th analysis, and $U$ is the decision boundary as described in section \ref{subsec32}.

Power is estimated as
\begin{align*}
    \text{Power} = \frac{\text{Number of times  correctly detecting the difference}}{\text{Number of simulation runs}}
\end{align*}
where correctly detecting the difference means that for some $k\leq K+1$, $$P_{\theta=\theta_0}(\hat{\theta}_k>\delta \vert D_k)>U$$
for $\theta_0$ the value of $\theta$ under the alternative hypothesis.

%\subsection{Continuous outcomes} \label{subsec41}
For continuous outcomes, %without loss of generality, suppose that the within-cluster variance $\sigma_W^2=1$ and the true population mean for control group $\mu_c=0$. 
the prior mean and variance for the population mean for both groups are fixed at 0 and 100, respectively. Various values of ICC were explored and the between-cluster variance $\sigma_B^2$ was determined through $\sigma_B^2=\frac{\sigma_W^2 \rho}{1-\rho}$.  To generate clustered continuous data, we first generate the $n$ cluster-specific means from normal distributions with mean $\mu_c$ (control group) and $\mu_t$ (treatment group) and variance $\sigma_B^2$. Then, within each cluster, $m$ samples are drawn from a normal distribution with mean equal to the cluster-specific mean and variance $\sigma_W^2$. The resulting $m~\times~n$ samples are expected to %be clustered continuous data and
satisfy the preset correlation structure. %The simulations parameters are summarized in Table \ref{tab:1}.
For binary outcomes, clustered binary data are generated via Beta and binomial distributions; see Appendix \ref{secA3} for details. All simulation parameters are summarized in Table \ref{tab:1}.

\begin{sidewaystable}[p]
%\notsotiny
\caption{Simulation parameters for continuous and binary outcome}
\begin{center}
\label{tab:1}
%\begin{adjustbox}{max width=\textwidth}
%\scriptsize
\normalsize
%\small
\begin{minipage}{\textwidth}
\begin{tabular}{lrrrr}
  \toprule
   & \multicolumn{2}{c}{Continuous} & \multicolumn{2}{c}{Binary} \\
  \cmidrule{2-3} \cmidrule{4-5}
  \multirow{2}{*}{Parameters}  & \multicolumn{1}{c}{Single}%Single interim analysis
  & \multicolumn{1}{c}{Multiple} %Multiple interim analyses & 
  & \multicolumn{1}{c}{Single}
  & \multicolumn{1}{c}{Multiple}
 \\
  &  \multicolumn{1}{c}{interim analysis} & \multicolumn{1}{c}{interim analyses} 
 & \multicolumn{1}{c}{interim analysis }
 & \multicolumn{1}{c}{interim analyses}
 \\
\midrule
  Population mean for control\footnote{\label{ft1}Not applicable to binary outcomes.}
  ($\mu_c$) & 0 & 0 & $\backslash$ & $\backslash$ \\
\addlinespace
 Within-cluster variance\footnoteref{ft1} 
 ($\sigma_W^2$) 
 & 1 & 1 & $\backslash$ & $\backslash$\\
 \addlinespace
 Baseline risk\footnote{Not applicable to continuous outcomes.} $(\pi_c$) & $\backslash$ & $\backslash$ & 0.25, 0.35, 0.45 & 0.25, 0.35, 0.45\\
\addlinespace
  Number of clusters per group ($n$) & 20, 40, 60 & 20, 40, 60 & 20, 40, 60& 20, 40, 60\\ 
  \addlinespace
   True treatment effect ($\Delta$) & 0, 0.1, \ldots, 0.9 & 0, 0.1, \ldots, 0.9 & 0, 0.1, 0.2, 0.3 & 0, 0.1, 0.2\\
  \addlinespace
   Decision boundary ($U$) & 0.95, 0.98 & 0.95, 0.98 & 0.95, 0.98 & 0.95, 0.98 \\
  \addlinespace
   Intra-cluster correlation coefficient ($\rho$) & 0.1, \ldots, 0.9 & 0.2, 0.5, 0.8 & 0.05, 0.1 & 0.05, 0.1 \\ 
  \addlinespace
   Cluster size ($m$)  & 8  & 8, 16 & 8 & 8, 16  \\ 
   \addlinespace
    Number of interim looks ($K$)  & 1 & 1, 2, 3 & 1& 1, 3\\
   \bottomrule
\end{tabular}
\end{minipage}
%\end{adjustbox}
\end{center}
\end{sidewaystable}

\begin{figure}[H]
     \centering
     \begin{subfigure}[b]{0.49\textwidth}
         \centering
         \includegraphics[width=\textwidth]{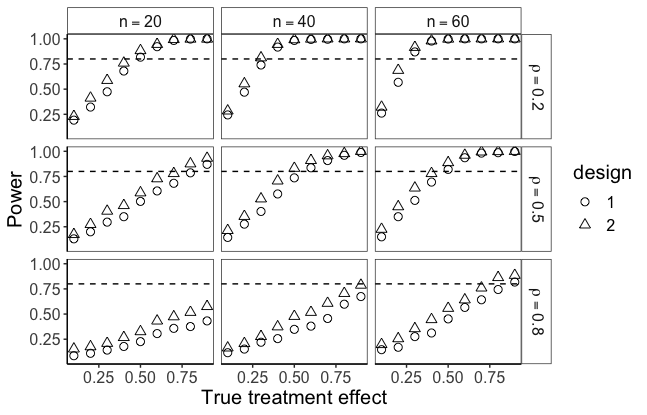}
         \caption{$U=0.95$}
         \label{PWR-continuous-single-U0.95}
     \end{subfigure}
     %\hfill
     \begin{subfigure}[b]{0.49\textwidth}
         \centering
         \includegraphics[width=\textwidth]{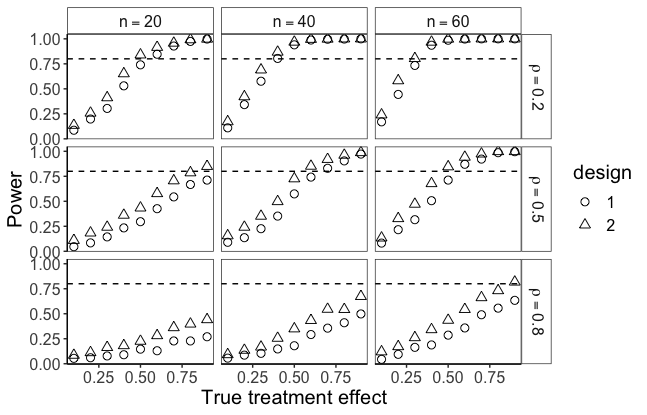}
         \caption{$U=0.98$}
         \label{PWR-continuous-single-U0.98}
     \end{subfigure}
        \caption{Plot of power versus true treatment effect when $n=20, 40, 60$, $\rho=0.2, 0.5, 0.8$ for (a) $U=0.95$ and (b) $U=0.98$ with single interim analysis planned. The dashed lines show the power of 0.8.}
        \label{PWR-continuous-single}
\end{figure}

Only results for continuous outcomes are presented here. The results for binary outcomes are summarized in Appendix \ref{secA3}. Figures \ref{FPR-continuous-single} and \ref{PWR-continuous-single} show the false positive rate and power when a single interim analysis is planned. For power, only results for $\rho=0.2, 0.5, 0.8$, representing low, moderate and high correlation, respectively, are displayed. In general, design 2 has higher power but also higher false positive rates compared with design 1. %, but it also has higher false positive rates. 

However, in all scenarios, with $U=0.95$,  the false positive rates for both designs are well above 0.05 and thus are unsatisfactory. With $U=0.98$, the false positive rates are reduced. Especially for design 1, the false positive rate can be controlled within 0.05 with $U=0.98$. Therefore, $U=0.98$ serves as a better decision boundary than $U=0.95$. A larger decision boundary corresponds to a more conservative test since more evidence is required to conclude the treatment efficacy. 

The effect of ICC on false positive rate is quite small but the effect on power is more direct. Under the same conditions, if the underlying ICC is higher, power is lower. For small ICC, if the underlying treatment effect is moderate to large, then only 20 clusters for each group will be sufficient to ensure the desired power, but if the underlying treatment effect is small, larger number of clusters, for example 40 or 60, should be considered. For moderate ICC and a small to moderate treatment effect, a larger number of clusters is recommended; for a large treatment effect only a small number of clusters, for example 20 clusters for each group may be enough. For large ICC, it is recommended to have more clusters than 60 as, even with a large treatment effect, power is still low. 

\begin{figure}[H]
     \centering
     \begin{subfigure}[b]{0.49\textwidth}
         \centering
         \includegraphics[width=\textwidth]{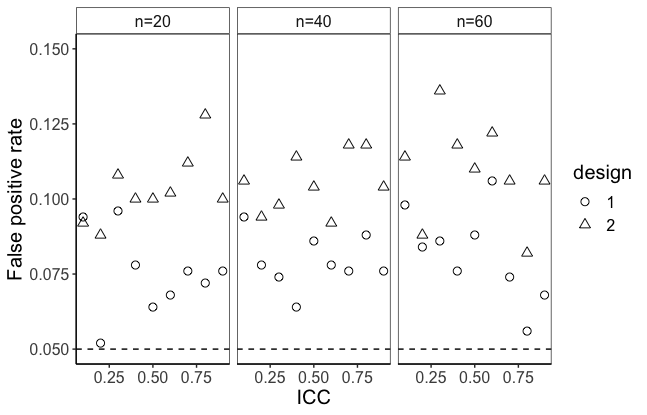}
         \caption{$U=0.95$}
         \label{FPR-continuous-single-U0.95}
     \end{subfigure}
     %\hfill
     \begin{subfigure}[b]{0.49\textwidth}
         \centering
         \includegraphics[width=\textwidth]{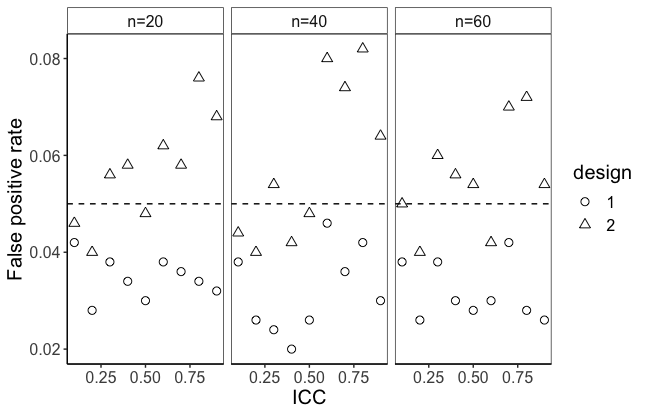}
         \caption{$U=0.98$}
         \label{FPR-continuous-single-U0.98}
     \end{subfigure}
        \caption{Plot of false positive rate versus ICC when $n=20, 40, 60$ for (a) $U=0.95$ and (b) $U=0.98$ with single interim analysis planned. The dashed lines show the false positive rate of 0.05. }
        \label{FPR-continuous-single}
\end{figure}

\begin{figure}[H]
     \centering
     \begin{subfigure}[b]{0.49\textwidth}
         \centering
         \includegraphics[width=\textwidth]{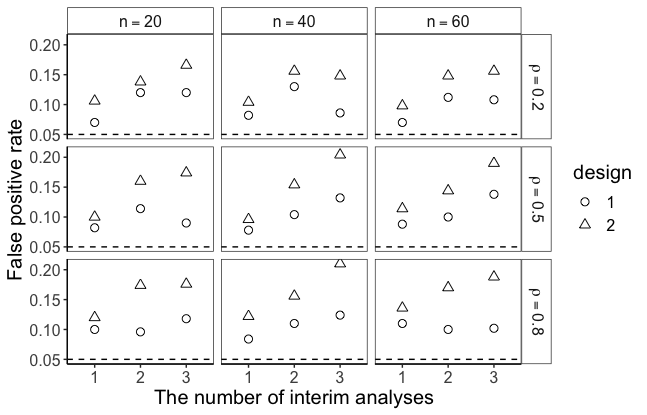}
         \caption{$U=0.95$}
         \label{FPR-continuous-multiple-m8-U0.95}
     \end{subfigure}
     %\hfill
     \begin{subfigure}[b]{0.49\textwidth}
         \centering
         \includegraphics[width=\textwidth]{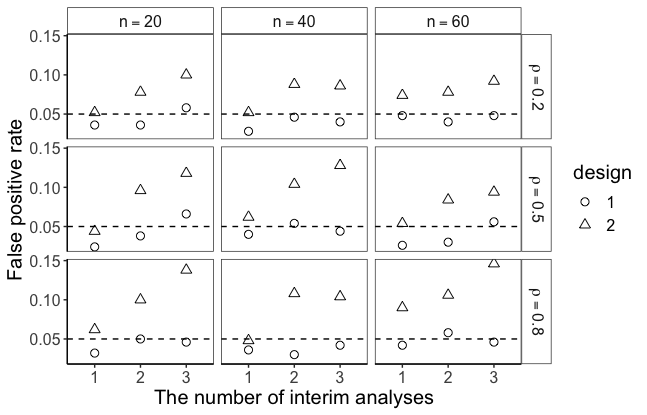}
         \caption{$U=0.98$}
         \label{FPR-continuous-multiple-m8-U0.98}
     \end{subfigure}
        \caption{Plot of false positive rate versus number of interim looks for $n=20, 40, 60$, $\rho = 0.2, 0.5, 0.8$, $m=8$ for (a) $U=0.95$ and (b) $U=0.98$. The dashed lines show the false positive rate of 0.05. }
        \label{FPR-continuous-multiple}
\end{figure}

\begin{figure}[H]
     \centering
     \begin{subfigure}[b]{0.49\textwidth}
         \centering
         \includegraphics[width=\textwidth]{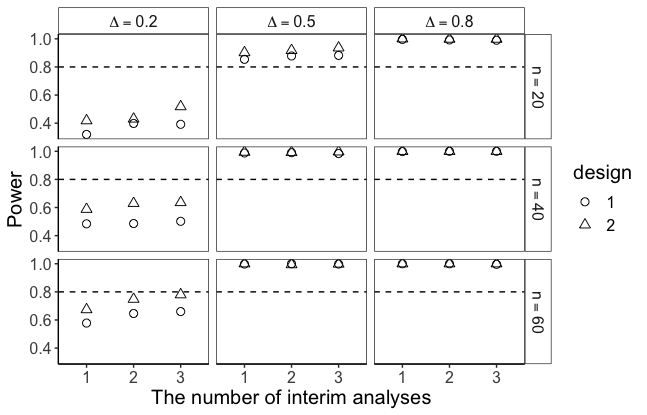}
         \caption{$\rho=0.2$, $U=0.95$}
         \label{PWR-continuous-multiple-rho0.2-m8-U0.95}
     \end{subfigure}
     %\hfill
     \begin{subfigure}[b]{0.49\textwidth}
         \centering
         \includegraphics[width=\textwidth]{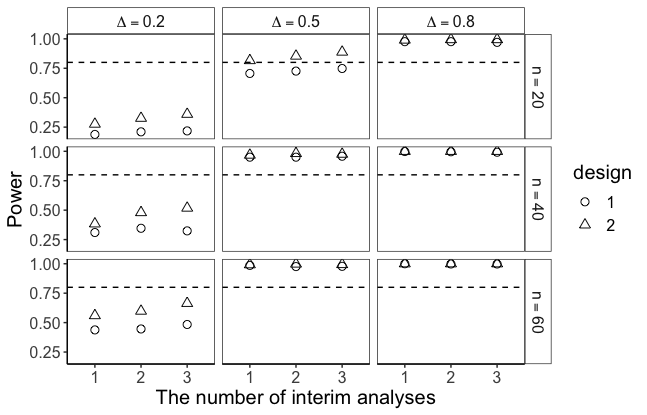}
         \caption{$\rho=0.2$, $U=0.98$}
         \label{PWR-continuous-multiple-rho0.2-m8-U0.98}
     \end{subfigure}\\
     \begin{subfigure}[b]{0.49\textwidth}
         \centering
         \includegraphics[width=\textwidth]{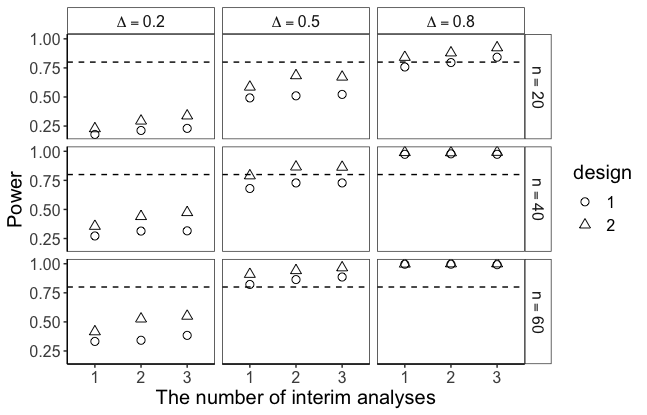}
         \caption{$\rho=0.5$, $U=0.95$}
         \label{PWR-continuous-multiple-rho0.5-m8-U0.95}
     \end{subfigure}
     %\hfill
     \begin{subfigure}[b]{0.49\textwidth}
         \centering
         \includegraphics[width=\textwidth]{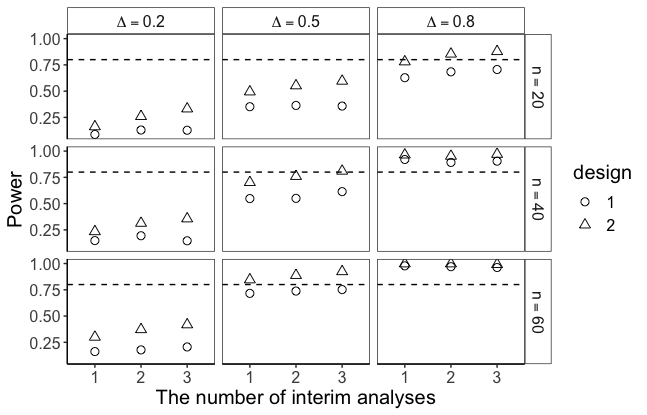}
         \caption{$\rho=0.5$, $U=0.98$}
         \label{PWR-continuous-multiple-rho0.5-m8-U0.98}
     \end{subfigure}\\
       \begin{subfigure}[b]{0.49\textwidth}
         \centering
         \includegraphics[width=\textwidth]{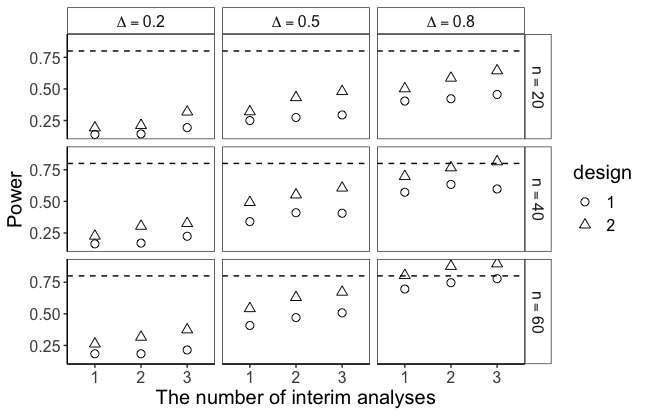}
         \caption{$\rho=0.8$, $U=0.95$}
         \label{PWR-continuous-multiple-rho0.8-m8-U0.95}
     \end{subfigure}
     %\hfill
     \begin{subfigure}[b]{0.49\textwidth}
         \centering
         \includegraphics[width=\textwidth]{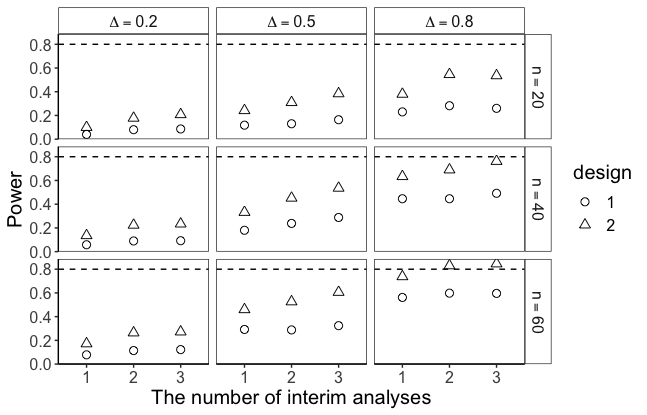}
         \caption{$\rho=0.8$, $U=0.98$}
         \label{PWR-continuous-multiple-rho0.8-m8-U0.98}
     \end{subfigure}
        \caption{Plot of power versus number of interim looks for $n=20, 40, 60$, $\Delta = 0.2, 0.5, 0.8$, $m=8$ with the subpanels (a)-(f) indicating all possible combinations of $\rho=0.2, 0.5, 0.8$ and $U=0.95, 0.98$. The dashed lines show the power of 0.8.}
        \label{PWR-continuous-multiple}
\end{figure}

Figures \ref{FPR-continuous-multiple} and \ref{PWR-continuous-multiple} show the false positive rate and power when multiple interim looks are built into the study design. Design 2 still has higher false positive rate and power compared with design 1. As was observed in Figures \ref{FPR-continuous-single} and \ref{PWR-continuous-single}, a larger decision boundary can reduce false positive rate and the resulting reduction in power can be remedied by recruiting a larger number of clusters. Adding more interim analyses may increase the false positive rate and the power. Given a fixed decision boundary, with more interim analyses planned, there is higher chance of rejecting the null hypothesis. %However, in some cases such as when $\Delta=0.8$, $\rho=0.2$ in Figures \ref{PWR-continuous-multiple-rho0.2-m8-U0.95} and \ref{PWR-continuous-multiple-rho0.2-m8-U0.98}, multiple interim analyses only result in negligible increase in power. The reason is that when the treatment effect is large and the ICC is relatively small, the power for a trial with only single interim analysis is already very high. There is little room for improvement.

As seen in the plot,  for large ICC values, the addition of multiple interim analyses can bring obvious improvement in power under design 2. However,% even if the power increases with the multiple interim looks, it is still unsatisfactory.
larger sample size is required for large ICC values to achieve sufficient power.
%Therefore, in design 2 it is preferable to recruit more clusters and only add multiple interim analyses when it is not feasible to increase the number of clusters.
For small to moderate ICC, regardless of design, a single interim analysis is preferred. The effect of cluster size on both false positive rate and power is very modest. The results for a larger cluster size are displayed in Figures \ref{FPR-continuous-multiple-m16} and \ref{PWR-continuous-multiple-m16} in Appendix \ref{secA2}.

To conclude, both designs perform better in terms of false positive rate with $U=0.98$, and design 2 has a higher false positive rate and power. The choice between the two designs as well as the two decision boundaries depends on the research question,  phase of the study, and the information we have about the treatment. For example, in some early phase clinical trials, if a relatively high false positive rate is acceptable, then design 2 with $U=0.95$ may be recommended. %since it has higher power. However, in late phase clinical trials, controlling false positive rate may be more important than maintaining a high power \cite{rct, type12}. Therefore, design 1 with a larger decision boundary such as $U=0.98$ may be recommended, since it is more conservative and the false positive rate can be controlled within 0.05.   
However, for the situation we explored, a single interim analysis is preferred no matter which design or decision boundary is chosen.

\section{Discussion} \label{sec5}
Motivated by a potential real-world cluster-randomized trial, we explore the statistical properties of Bayesian adaptive cluster-randomized trial designs. We explore stopping rules for efficacy in a cluster-randomized trial. 
Interim analyses may be planned over the course of the trial and at each interim analysis the trial may be stopped early if sufficient evidence is established to conclude efficacy otherwise the trial proceeds to the next analysis. We proposed two designs which sequentially enroll participants in different ways. The first design sequentially enrolls clusters, and individual participants for each cluster are recruited all together. The second recruits all clusters at the start of the trial, then sequentially enrolls batches of participants. % at the subsequent analysis point.
Regardless of the design choice, the data analysis procedure is the same: %will be the same since 
the difference between the designs lies in the sequential enrollment of participants. 

Interim analyses of both continuous and binary outcomes are performed based on common models. For continuous outcomes, on the basis of the normality assumption, we obtained the analytical form of the posterior distribution of the population mean. Based on Monte Carlo simulation, the posterior probability of efficacy can be easily estimated by drawing random samples directly from the two posterior distributions, one for each treatment group. For binary outcomes, due to the complex correlation structure, the posterior distribution of population proportion cannot be obtained analytically. Instead, a hierarchical model  was used and the posterior probability of efficacy is estimated via MCMC. % implemented in \texttt{RStan}. 

%Simulation studies are carried out for a wide variety of scenarios. 
Through simulation, we found the design choice may be different for different outcomes. The preferred design may also depend on the research goal, phase of study, or feasibility considerations. %For example,   
%in some early phase clinical trials, design 2 may be recommended since it can maintain higher power and a relatively high false positive rate may be acceptable in early phase clinical trials. However, in late
 %in late phase clinical trials, controlling false positive rate may be more important to clinical scientists, while in early phase clinical trials, a relatively high false positive rate may be acceptable as the goal is to identify potentially useful interventions. Therefore, the design choice as well as the choice of decision boundary for continuous outcome in early and late phase clinical trials can be different. 
For binary outcomes, within the parameter space we explored, design 2 is recommended based on design operating characteristics. Also, for both outcomes, one interim analysis may be recommended for the situations we investigated, as it makes the designs more feasible and the design performance with single interim analysis is satisfactory. However, a more general recommendation requires further exploration to account for practical issues or other design parameters we did not explore in our simulation. In addition, increasing cluster size may not necessarily bring much improvement in design performance. Keeping a small cluster size may not hurt design operating characteristics very much but it will save time and cost. In our motivating trial, a smaller cluster size is also more realistic, as it reduces the burden on the individual clinician to find suitable patients from their practice.

There are some limitations to our work. % and we expect to improve them in the future.
First, we only considered implementing a stopping rule for efficacy. There are many other adaptive features that we have not considered. One example is adaptive randomization. The response-adaptive randomization \cite{discussionrar} has not been extensively investigated from Bayesian viewpoint. %In the future it may be worth investigating, as it would enable adjusting allocation ratio based on interim data, or more specifically, the posterior probability and distribution so that more subjects are allocated to the most appropriate treatment. 
Second, we only focus on two-arm trials. Extension to multi-arm trials which may involve arm dropping can be considered in the future. Third, we only considered continuous and binary outcomes. However, survival endpoints are also prevalent in clinical trials. An extension to survival outcomes based on survival models can also improve the framework of Bayesian adaptive cluster-randomized trial. In addition, the effect of unequal cluster sizes has not been explored for our designs. However,  the effect of varying cluster sizes on design operating characteristics in standard cluster-randomized trials has been extensively discussed \cite{crtintro7, discussunequalcls1, discussunequalcls2, discussunequalcls3, discussunequalcls4}. Typically, designs are most efficient for equal cluster sizes. Inflation of false positive rate will occur for imbalanced studies, and with the same sample size, imbalanced trials may be underpowered compared with their balanced counterparts. Similar impacts of unequal cluster sizes on power or false positive rate may also exist for our Bayesian adaptive cluster-randomized trials, and this would need to be determined in future work. Also, practical issues in planning a Bayesian adaptive cluster-randomized trial are not taken into account in our paper. For example, sometimes it may not be realistic to plan the preferred design recommended by design operating characteristics. In this case, feasibility may be the main reason for design choice. These context-specific issues require more exploration in the future.

\section*{Declaration of competing interest}
D.B. is a director, shareholder, and employee of Aifred Health.

\section*{Acknowledgements}
This work was supported by funding from MITACS Accelerate Grant \#ACC IT18791. %SG acknowledges funding via a Discovery Grant from Natural Sciences and Engineering Research Council of Canada (NSERC). EEMM is a Canada Research Chair (Tier 1) in Statistical Methods for Precision Medicine and  acknowledges the support of a chercheur de mérite award from the Fonds de recherche du Québec-Santé. 

\bibliography{sn-article.bib}{}

\newpage
\begin{appendices}

\section{Propositions}\label{secA1}

\begin{proposition}
\label{lemma1}
For $Y_{ij}$ satisfying $Y_{ij}\vert \mu_j \sim N(\mu_j, \sigma_W^2)$ and $\mu_j\sim N(\mu, \sigma_B^2)$, the marginal distribution of $Y_{ij}$ is normal with mean $\mu$ and variance $\sigma_W^2+\sigma_B^2$.
\end{proposition}

\begin{proof}
The marginal distribution of $Y_{ij}$ can be obtained from 
\begin{align*}
    f(Y_{ij})&=\int_{-\infty}^{+\infty} f(Y_{ij},\mu_j) d \mu_j\\
        &=\int_{-\infty}^{+\infty}f(Y_{ij}\vert\mu_j)f(\mu_j) d \mu_j\\
        &=\int_{-\infty}^{+\infty} \frac{1}{\sqrt{2\pi \sigma_W^2}}e^{-\frac{(Y_{ij}-\mu_j)^2}{2\sigma_W^2}}\times \frac{1}{\sqrt{2\pi \sigma_B^2}}e^{-\frac{(\mu_j-\mu)^2}{2\sigma_B^2}} d \mu_j\\
        &=\frac{1}{2\pi\sqrt{\sigma_W^2\sigma_B^2}} \text{exp}\left\{-\frac{Y_{ij}^2}{2\sigma_W^2}-\frac{\mu^2}{2\sigma_B^2}\right\} \times \\
        &\int_{-\infty}^{+\infty} \text{exp}\left\{-\frac{1}{2}\left (\frac{1}{\sigma_B^2}+\frac{1}{\sigma_W^2}\right)\mu_j^2+\mu_j\left (\frac{Y_{ij}}{\sigma_W^2}+\frac{\mu}{\sigma_B^2}\right)\right\} d\mu_j .
\end{align*}
We can calculate the integral using the pdf of normal distribution,
\begin{align*}
    &\int_{-\infty}^{+\infty} \text{exp}\left\{-\frac{1}{2}\left (\frac{1}{\sigma_B^2}+\frac{1}{\sigma_W^2}\right)\mu_j^2+\mu_j\left (\frac{Y_{ij}}{\sigma_W^2}+\frac{\mu}{\sigma_B^2}\right)\right\} d\mu_j\\ &=\text{exp}\left\{\frac{\left(Y_{ij}\sigma_W^{-2}+\mu\sigma_B^{-2}\right)^2}{2\left (\sigma_W^{-2}+\sigma_B^{-2}\right)}\right\} \times \int_{-\infty}^{+\infty} \text{exp}\left\{-\frac{\left(\mu_j-\frac{Y_{ij}\sigma_W^{-2}+\mu \sigma_B^{-2}}{\sigma_B^{-2}+\sigma_W^{-2}}\right)^2}{2\left(\sigma_B^{-2}+\sigma_W^{-2}\right)^{-1}}\right\} d \mu_j\\
    &=\sqrt{2\pi} \left(\sigma_B^{-2}+\sigma_W^{-2}\right)^{-\frac{1}{2}}\text{exp}\left\{\frac{\left(  Y_{ij}\sigma_W^{-2}+\mu \sigma_B^{-2} \right)^2}{2 \left(\sigma_B^{-2}+\sigma_W^{-2} \right)}  \right\} .
\end{align*}
Therefore, the marginal distribution is
\begin{align*}
        f(Y_{ij})&=\frac{1}{\sqrt{2\pi \left( \sigma_B^2+\sigma_W^2 \right)}} \text{exp}\left\{ -\frac{Y_{ij}^2}{2\sigma_W^2}-\frac{\mu^2}{2\sigma_B^2}+\frac{\left(  Y_{ij}\sigma_W^{-2}+\mu \sigma_B^{-2} \right)^2}{2 \left(\sigma_B^{-2}+\sigma_W^{-2} \right)}  \right\}\\
        &=\frac{1}{\sqrt{2\pi \left( \sigma_B^2+\sigma_W^2 \right)}}\text{exp}\left\{-\frac{(Y_{ij}-\mu)^2}{2(\sigma_B^2+\sigma_W^2)} \right\} .
\end{align*}
That is, $Y_{ij} \sim N(\mu, \sigma_B^2+\sigma_W^2)$.
\end{proof}

\newpage
\begin{proposition}
\label{lemma2}
The response vector $Y=(Y_1^T, \cdots, Y_{n_k}^T)^T$ where $Y_j=(Y_{1j},\cdots,Y_{m_kj})^T$, $j=1, \cdots, n_k$, and $Y_{ij}$ satisfies the conditions stated in proposition \ref{lemma1},  follows a multivariate normal distribution with mean $\mathbf{\overrightarrow{\mu}}=(\mu, \ldots, \mu)_{m_k\times n_k}$ and variance-covariance matrix $\Sigma$ as a block matrix of the form 
$$\begin{pmatrix}
\Sigma_{Y_1} & \mathbf{0}  &\cdots & \mathbf{0}\\
\mathbf{0} & \Sigma_{Y_2}  & \cdots &\mathbf{0}\\
\vdots & \vdots & \ddots  &\mathbf{0}\\
\mathbf{0} & \mathbf{0} & \cdots &\Sigma_{Y_{n_k}}
\end{pmatrix}
$$
where $\Sigma_{Y_j}=\text{Cov}(Y_j)$ satisfies 

\begin{align*}
        \text{Var}(Y_{ij})&=\sigma_B^2+\sigma_W^2\\
        \text{Cov}(Y_{ij}, Y_{sj})&=\sigma_B^2, \quad i\neq s\\
        \text{Cov}(Y_{ij}, Y_{tl})&=0, \quad j\neq l
\end{align*}
%$\text{Var}(Y_{ij})=\sigma_B^2+\sigma_W^2$, $\text{Cov}(Y_{ij}, Y_{kj})=\sigma_B^2$, $i\neq k$ and $\text{Cov}(Y_{ij}, Y_{kl})=0$, $j\neq l$.
\end{proposition}

\begin{proof}
We only need to prove that $\forall$ $(a_{11}, a_{21}, \cdots, a_{mn})$, $Z=a_{11}Y_{11}+\cdots+a_{mn}Y_{mn}$ is univariate normal. Let $M_Z(t)$ be the moment generating function of $Z$. Therefore, 
\begin{align*}
    \begin{split}
        M_Z(t)&=E\left(\text{exp}\{tZ\}\right)\\
        &=E\left(\text{exp}\left\{t\sum_{j=1}^m \sum_{i=1}^n a_{ij}Y_{ij}\right \}\right)\\
        &=E\left [E\left( \text{exp} \left\{ t\sum_{j=1}^n\sum_{i=1}^m a_{ij}Y_{ij}\right\} \vert \mu_1,\ldots,\mu_j\right)\right]\\
        &=E\left[\prod_{j=1}^n E\left[\text{exp}\left\{t\sum_{i=1}^m a_{ij} Y_{ij}\right\}\vert \mu_j\right]\right]\\
       &=E\left[ \prod_{j=1}^n \prod_{i=1}^m E\left(\text{exp}\{ta_{ij}Y_{ij}\}\vert \mu_j\right)\right ]\\
        &=E\left [ \prod_{j=1}^n \prod_{i=1}^m \text{exp}\left\{ ta_{ij}\mu_j+\sigma_W^2a_{ij}^2 t^2/2 \right\}\right]\\
        &=\text{exp} \left\{ \sigma_W^2t^2 (\sum_{j=1}^n \sum_{i=1}^m a_{ij}^2)/2\right \}\prod_{j=1}^n E\left [\text{exp}\left\{t\mu_j \sum_{i=1}^m a_{ij}\right\}\right]\\
        &=\text{exp} \left \{  t\mu \sum_{j=1}^{n} \sum_{i=1}^m a_{ij} +\frac{t^2}{2}\left[ \sigma_W^2\left(  \sum_{j=1}^{n}\sum_{i=1}^m a_{ij}^2\right)+\sigma_B^2\sum_{j=1}^n \left ( \sum_{i=1}^m a_{ij} \right)^2 \right] \right \} .
    \end{split}
\end{align*}
This is the moment generating function of a normal random variable. Therefore, $Z$ is univariate normal and $Y$ is multivariate normal. Then the mean of $Y$ is $\mathbf{\overrightarrow{\mu}}=(\mu,\ldots,\mu)^T$ and the variance-covariance structure satisfies
\begin{align*}
    \text{Var}(Y_{ij})&=\sigma_B^2+\sigma_W^2\\
    \text{Cov}(Y_{ij}, Y_{kj})&=E(Y_{ij}Y_{kj})-E(Y_{ij})E(Y_{kj})\\
    &=E[E(Y_{ij}Y_{kj}\vert\mu_j)]-\mu^2\\
    &=E(\mu_j^2)-\mu^2\\
    &=\sigma_B^2\\
    \text{Cov}(Y_{ij}, Y_{kl})&=E(Y_{ij}Y_{kl})-E(Y_{ij})E(Y_{kl})\\
    &=E[E(Y_{ij}Y_{kl}\vert \mu_j, \mu_l)]-\mu^2\\
    &=\mu^2-\mu^2\\
    &=0.
\end{align*}
\end{proof}

\begin{proposition}
Assume a prior distribution $N(a,b^2)$ for $\mu$, under the conditions in proposition \ref{lemma1} and proposition \ref{lemma2},  the posterior distribution of $\mu$ is 
$$\mu\vert Y \sim N\left(\frac{b^2\textbf{1}^T\Sigma^{-1}Y+a}{b^2\textbf{1}^T\Sigma^{-1}\textbf{1}+1},\frac{b^2}{b^2\textbf{1}^T\Sigma^{-1}\textbf{1}+1}\right).$$
\end{proposition}

\begin{proof}
From proposition \ref{lemma2}, the response vector Y $\sim$ MVN($\overrightarrow{\mu}$, $\Sigma$). Therefore, we have

\begin{align*}
    f(\mu\vert Y)&\propto f(Y \vert \mu)f(\mu)\\
    &=(2\pi)^{\frac{mn}{2}} \vert \Sigma \vert^{-\frac{1}{2}}\exp \left\{ -\frac{1}{2}(Y-\mu \mathbf{1})^T \Sigma^{-1}(Y-\mu \mathbf{1})\right\} \times \frac{1}{\sqrt{2\pi b^2}}\exp \left\{-\frac{(\mu-a)^2}{2b^2}\right\}\\
    &\propto \exp \left\{-\frac{1}{2}(Y-\mu \mathbf{1})^T \Sigma^{-1}(Y-\mu \mathbf{1})- \frac{(\mu-a)^2}{2b^2} \right\}\\
    &\propto \exp \left \{ -\frac{\mu^2}{2} \frac{b^2 \mathbf{1}^T\Sigma^{-1}\mathbf{1}+1}{b^2}+\frac{b^2\mathbf{1}^T\Sigma^{-1}Y+a}{b^2}\mu \right\}\\
    &\propto \exp \left \{ -\frac{b^2\mathbf{1}^T \Sigma^{-1}\mathbf{1}+1}{2b^2}\left(\mu-\frac{b^2\mathbf{1}^T\Sigma^{-1}Y+a}{b^2\mathbf{1}^T\Sigma^{-1}\mathbf{1}+1}\right)^2\right \}.
\end{align*}
That is, 
\begin{equation*}
    \mu \vert Y \sim N\left(\frac{b^2\textbf{1}^T\Sigma^{-1}Y+a}{b^2\textbf{1}^T\Sigma^{-1}\textbf{1}+1},\frac{b^2}{b^2\textbf{1}^T\Sigma^{-1}\textbf{1}+1}\right).
    \end{equation*}
\end{proof}

\section{Results for continuous outcomes with cluster size m=16}
\label{secA2}

\begin{figure}[H]
     \centering
     \begin{subfigure}[b]{0.49\textwidth}
         \centering
         \includegraphics[width=\textwidth]{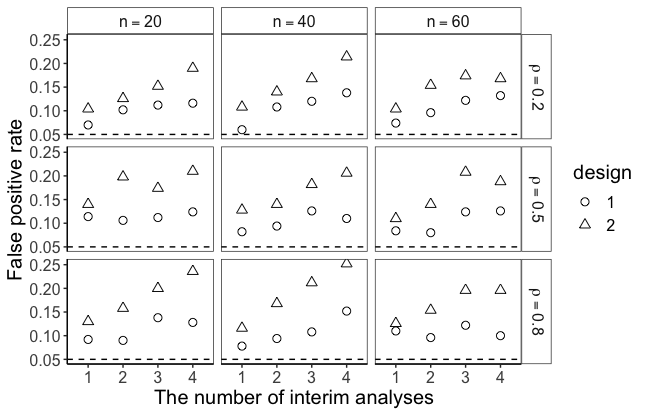}
         \caption{$U=0.95$}
         \label{PWR-continuous-multiple-m16-U0.95}
     \end{subfigure}
     %\hfill
     \begin{subfigure}[b]{0.49\textwidth}
         \centering
         \includegraphics[width=\textwidth]{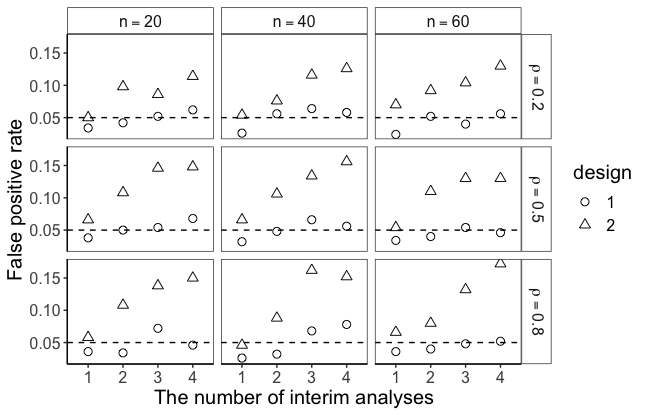}
         \caption{$U=0.98$}
         \label{FPR-continuous-multiple-m16-U0.98}
     \end{subfigure}
        \caption{Plot of false positive rate versus number of interim looks for $n=20, 40, 60$, $\rho = 0.2, 0.5, 0.8$, $m=16$ for (a) $U=0.95$ and (b) $U=0.98$. The dashed lines show the false positive rate of 0.05.}
        \label{FPR-continuous-multiple-m16}
\end{figure}

\begin{figure}[H]
     \centering
     \begin{subfigure}[b]{0.49\textwidth}
         \centering
         \includegraphics[width=\textwidth]{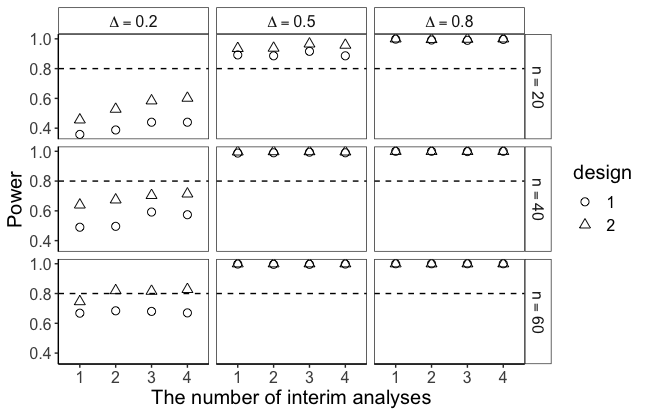}
         \caption{$\rho=0.2$, $U=0.95$}
         \label{PWR-continuous-multiple-rho0.2-m16-U0.95}
     \end{subfigure}
     %\hfill
     \begin{subfigure}[b]{0.49\textwidth}
         \centering
         \includegraphics[width=\textwidth]{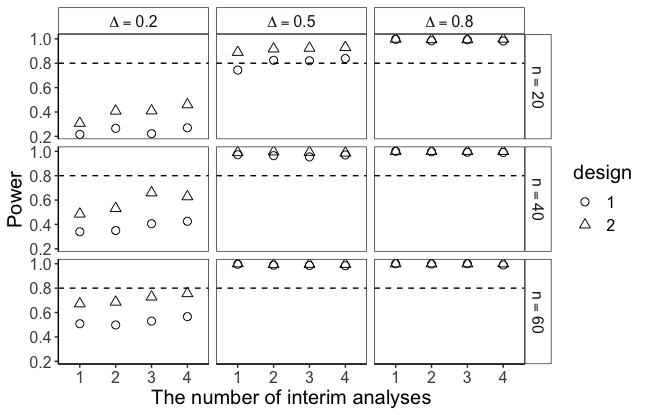}
         \caption{$\rho=0.2$, $U=0.98$}
         \label{PWR-continuous-multiple-rho0.2-m16-U0.98}
     \end{subfigure}\\
     \begin{subfigure}[b]{0.49\textwidth}
         \centering
         \includegraphics[width=\textwidth]{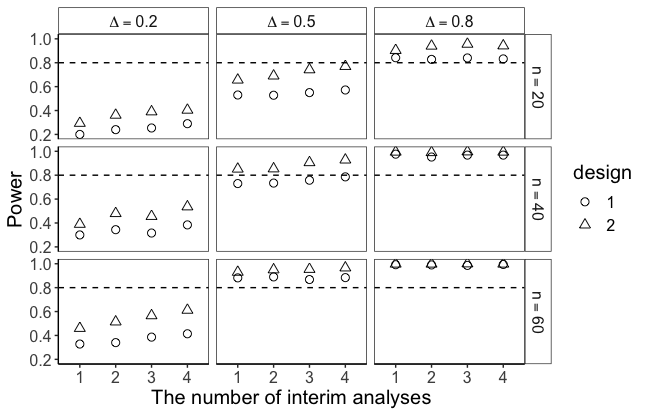}
         \caption{$\rho=0.5$, $U=0.95$}
         \label{PWR-continuous-multiple-rho0.5-m16-U0.95}
     \end{subfigure}
     %\hfill
     \begin{subfigure}[b]{0.49\textwidth}
         \centering
         \includegraphics[width=\textwidth]{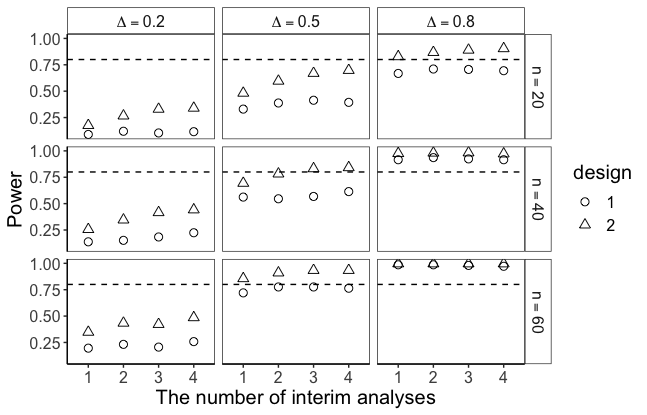}
         \caption{$\rho=0.5$, $U=0.98$}
         \label{PWR-continuous-multiple-rho0.5-m16-U0.98}
     \end{subfigure}\\
       \begin{subfigure}[b]{0.49\textwidth}
         \centering
         \includegraphics[width=\textwidth]{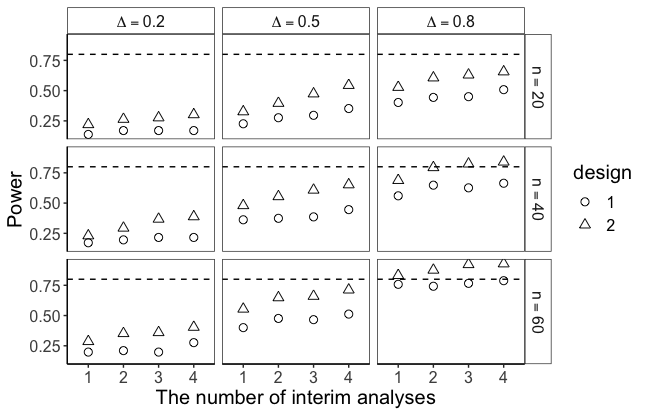}
         \caption{$\rho=0.8$, $U=0.95$}
         \label{PWR-continuous-multiple-rho0.8-m16-U0.95}
     \end{subfigure}
     %\hfill
     \begin{subfigure}[b]{0.49\textwidth}
         \centering
         \includegraphics[width=\textwidth]{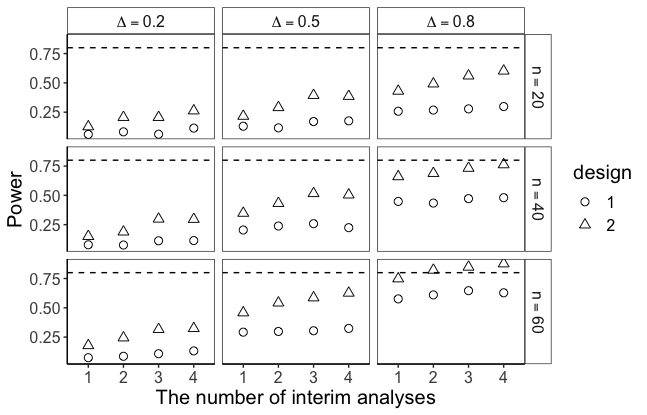}
         \caption{$\rho=0.8$, $U=0.98$}
         \label{PWR-continuous-multiple-rho0.8-m16-U0.98}
     \end{subfigure}
        \caption{Plot of power versus number of interim looks for $n=20, 40, 60$, $\Delta = 0.2, 0.5, 0.8$,$m=16$ with subpanels (a)-(f) indicating all possible combinations of $\rho = 0.2, 0.5, 0.8$ and $U=0.95, 0.98$. The dashed lines show the power of 0.8.}
        \label{PWR-continuous-multiple-m16}
\end{figure}

\newpage

\section{Simulation for binary outcomes}
\label{secA3}

To generate clustered binary data, we first generate $n$ cluster-specific proportions from Beta distributions with Beta parameters determined by the preset population mean and ICC so that all cluster-specific proportions are strictly between 0 and 1. Then, within each cluster, the number of events were generated from a binomial distribution with the cluster size at current stage and the cluster-specific proportions. Then the resulting $m\times n$ binary variables satisfy the predetermined correlation structure.

\begin{figure}[H]
     \centering
     \begin{subfigure}[b]{0.49\textwidth}
         \centering
         \includegraphics[width=\textwidth]{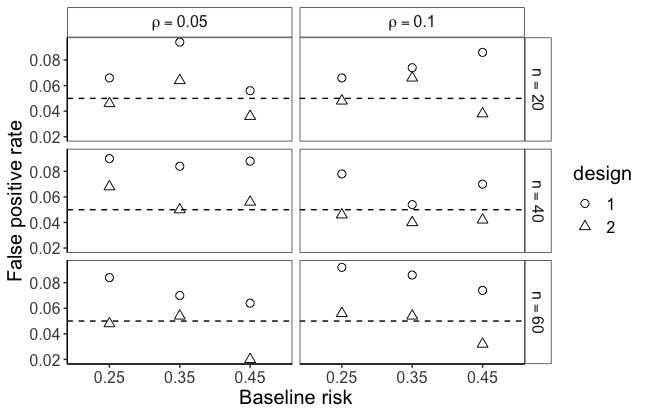}
         \caption{$U=0.95$}
         \label{FPR-binary-single-U0.95}
     \end{subfigure}
     %\hfill
     \begin{subfigure}[b]{0.49\textwidth}
         \centering
         \includegraphics[width=\textwidth]{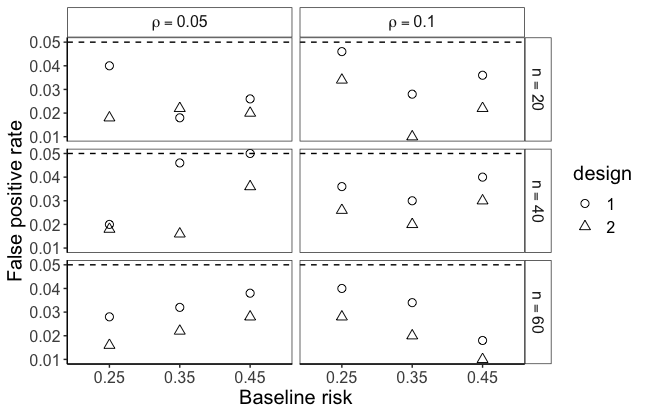}
         \caption{$U=0.98$}
         \label{FPR-binary-single-U0.98}
     \end{subfigure}
        \caption{Plot of false positive rate versus baseline risk when $n=20, 40, 60$, $\rho = 0.05, 0.1$ for (a) $U=0.95$ and (b) $U=0.98$ with single interim analysis planned. The dashed lines show the false positive rate of 0.05.}
        \label{FPR-binary-single}
\end{figure}

\begin{figure}[H]
     \centering
     \begin{subfigure}[b]{0.49\textwidth}
         \centering
         \includegraphics[width=\textwidth]{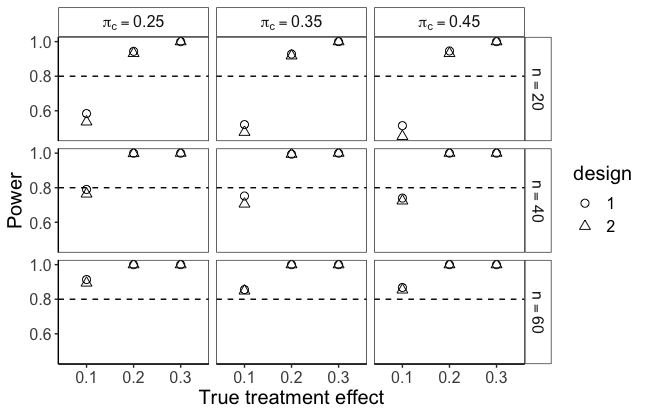}
         \caption{$\rho=0.05$, $U=0.95$}
         \label{PWR-binary-single-rho0.05-U0.95}
     \end{subfigure}
     %\hfill
     \begin{subfigure}[b]{0.49\textwidth}
         \centering
         \includegraphics[width=\textwidth]{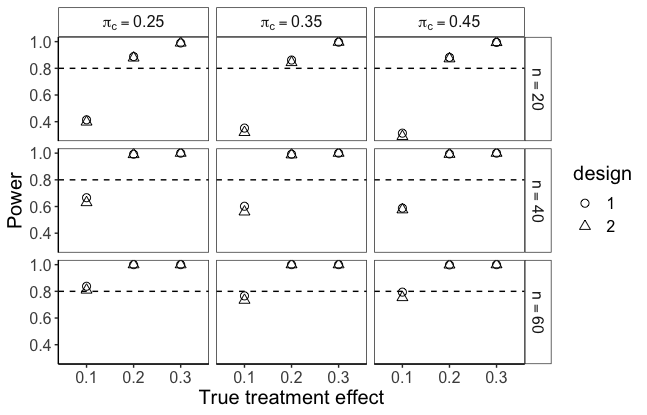}
         \caption{$\rho=0.05$, $U=0.98$}
         \label{PWR-binary-single-rho0.05-U0.98}
     \end{subfigure}\\
     \begin{subfigure}[b]{0.49\textwidth}
         \centering
         \includegraphics[width=\textwidth]{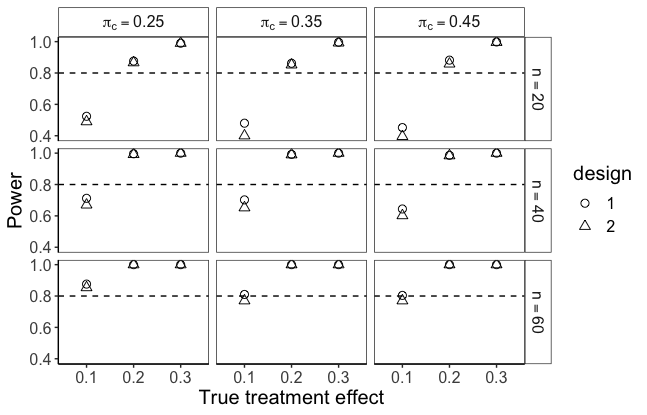}
         \caption{$\rho=0.1$, $U=0.95$}
         \label{PWR-binary-single-rho0.1-U0.95}
     \end{subfigure}
     %\hfill
     \begin{subfigure}[b]{0.49\textwidth}
         \centering
         \includegraphics[width=\textwidth]{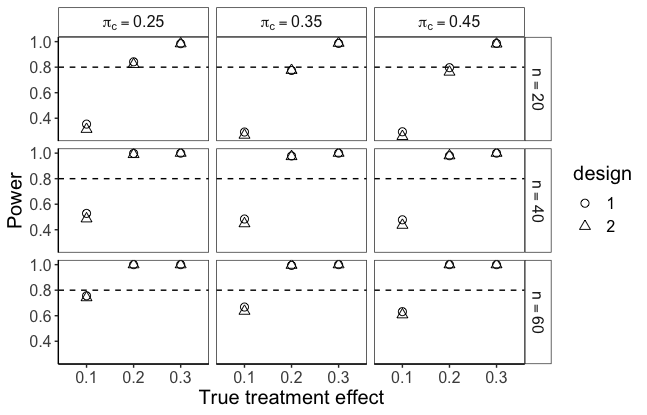}
         \caption{$\rho=0.1$, $U=0.98$}
         \label{PWR-binary-single-rho0.1-U0.98}
     \end{subfigure}
        \caption{Plot of power versus true treatment effect for $n=20, 40, 60$, $\pi_c = 0.25,0.35, 0.45$, with the subpanels (a)-(d) indicating all possible combinations of $\rho = 0.05, 0.1$ and $U=0.95, 0.98$ for single interim analysis. The dashed lines show the power of 0.85.}
        \label{PWR-binary-single}
\end{figure}

The performance of the two designs in terms of false positive rate and power, when only one interim analysis is planned, is displayed in Figures \ref{FPR-binary-single} and \ref{PWR-binary-single}. It can be observed that design 1 has higher false positive rate and both designs perform almost equally well in terms of power over the parameter space we explored. Also, a decision boundary of 0.98 helps control false positive rate compared with $U=0.95$ while the decrease in power coming with a larger decision boundary is acceptable. However, for design 2, even with a smaller decision boundary, the false positive rates for most cases are acceptable considering that they are fluctuating around 0.05. In addition, as in the continuous case, when the underlying ICC is larger,  power is relatively small. 

\begin{figure}[H]
     \centering
     \begin{subfigure}[b]{0.49\textwidth}
         \centering
         \includegraphics[width=\textwidth]{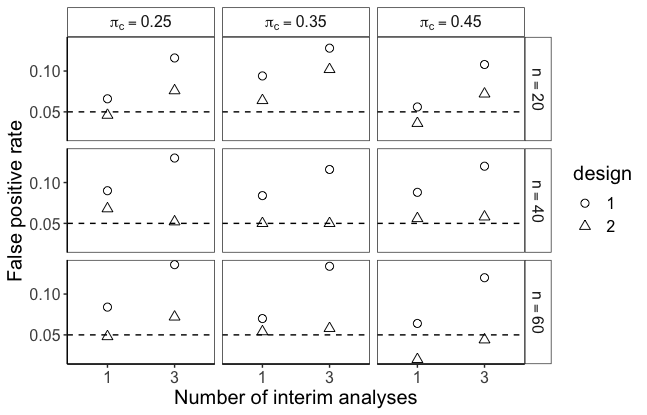}
         \caption{$\rho=0.05$, $U=0.95$}
         \label{FPR-binary-multiple-rho0.05-m8-U0.95}
     \end{subfigure}
     %\hfill
     \begin{subfigure}[b]{0.49\textwidth}
         \centering
         \includegraphics[width=\textwidth]{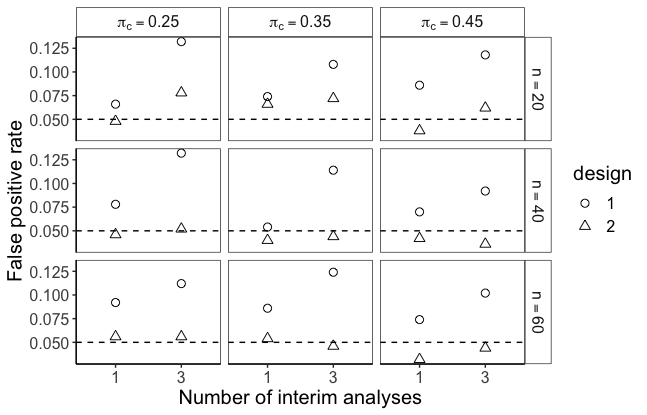}
         \caption{$\rho=0.1$, $U=0.95$}
         \label{FPR-binary-multiple-rho0.1-m8-U0.95}
     \end{subfigure}\\
     \begin{subfigure}[b]{0.49\textwidth}
         \centering
         \includegraphics[width=\textwidth]{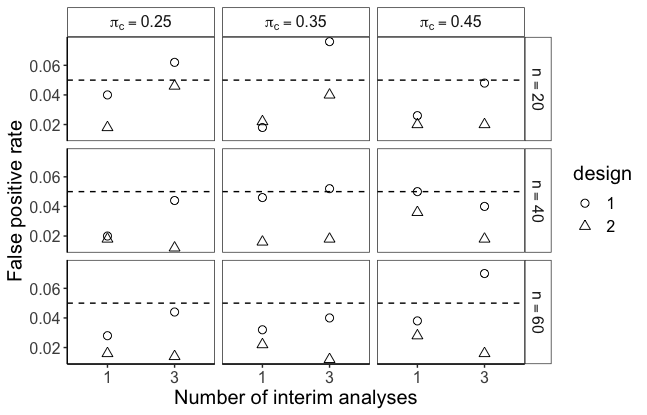}
         \caption{$\rho=0.05$, $U=0.98$}
         \label{FPR-binary-multiple-delta0.1-rho0.05-m8-U0.98}
     \end{subfigure}
     %\hfill
     \begin{subfigure}[b]{0.49\textwidth}
         \centering
         \includegraphics[width=\textwidth]{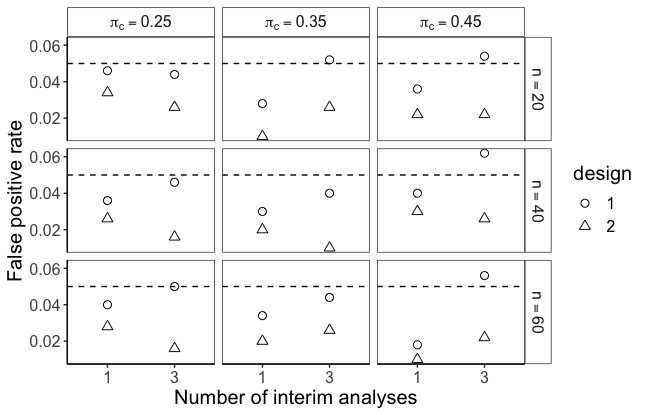}
         \caption{$\rho=0.1$, $U=0.98$}
         \label{FPR-binary-multiple-delta0.1-rho0.1-m8-U0.98}
     \end{subfigure}\\
        \caption{Plot of false positive rate versus number of interim looks for $n=20, 40, 60$, $\pi_c = 0.25, 0.35, 0.45$, $m=8$ with subpanels (a)-(d) indicating all possible combinations of $\rho = 0.05, 0.1$ and $U=0.95, 0.98$. The dashed lines show the false positive rate of 0.05.}
        \label{FPR-binary-multiple}
\end{figure}

\begin{figure}[H]
     \centering
     \begin{subfigure}[b]{0.49\textwidth}
         \centering
         \includegraphics[width=\textwidth]{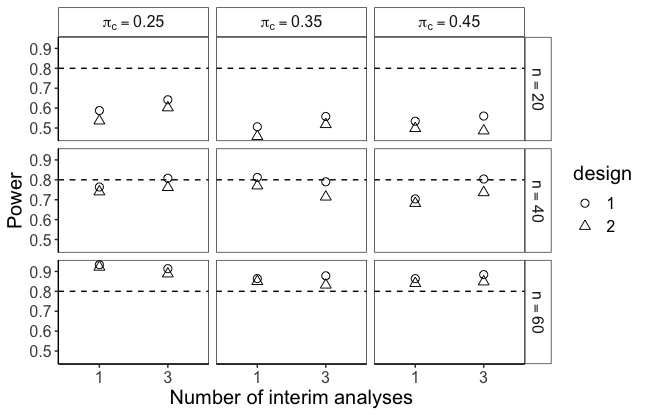}
         \caption{$\rho=0.05$, $U=0.95$, $\Delta=0.1$}
         \label{PWR-binary-multiple-delta0.1-rho0.05-m8-U0.95}
     \end{subfigure}
     %\hfill
     \begin{subfigure}[b]{0.49\textwidth}
         \centering
         \includegraphics[width=\textwidth]{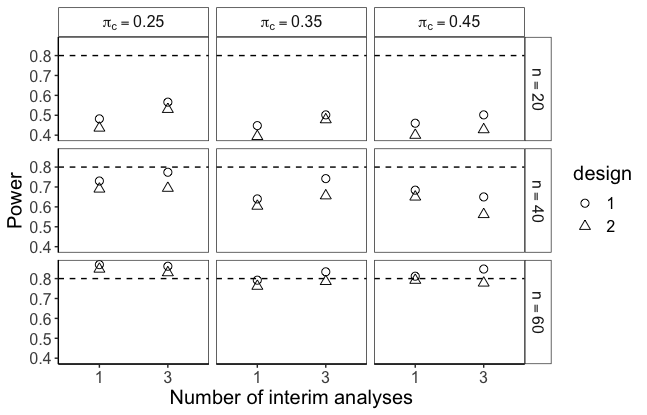}
         \caption{$\rho=0.1$, $U=0.95$, $\Delta=0.1$}
         \label{PWR-binary-multiple-delta0.1-rho0.1-m8-U0.95}
     \end{subfigure}\\
     \begin{subfigure}[b]{0.49\textwidth}
         \centering
         \includegraphics[width=\textwidth]{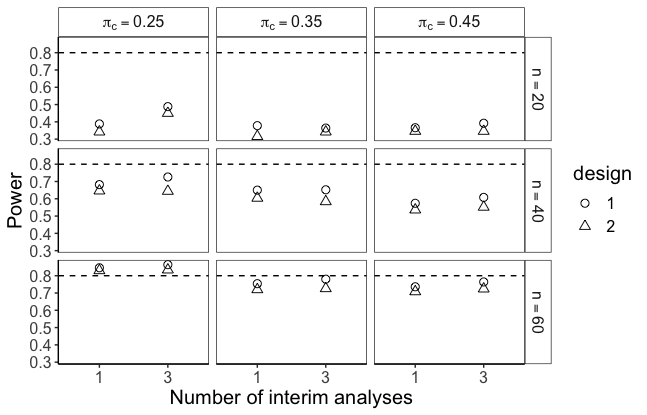}
         \caption{$\rho=0.05$, $U=0.98$, $\Delta=0.1$}
         \label{PWR-binary-multiple-delta0.1-rho0.05-m8-U0.98}
     \end{subfigure}
     %\hfill
     \begin{subfigure}[b]{0.49\textwidth}
         \centering
         \includegraphics[width=\textwidth]{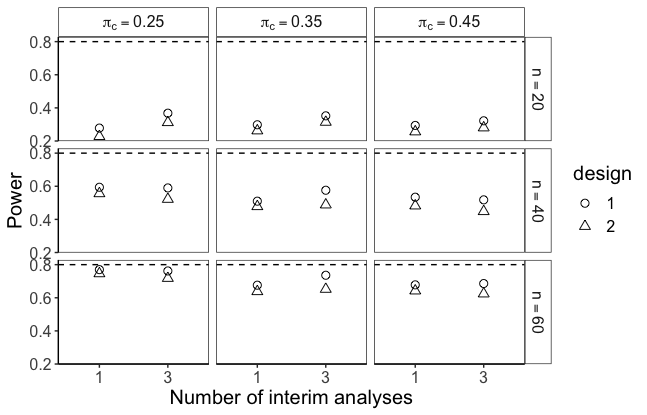}
         \caption{$\rho=0.1$, $U=0.98$, $\Delta=0.1$}
         \label{PWR-binary-multiple-delta0.1-rho0.1-m8-U0.98}
     \end{subfigure}\\
      \begin{subfigure}[b]{0.49\textwidth}
         \centering
         \includegraphics[width=\textwidth]{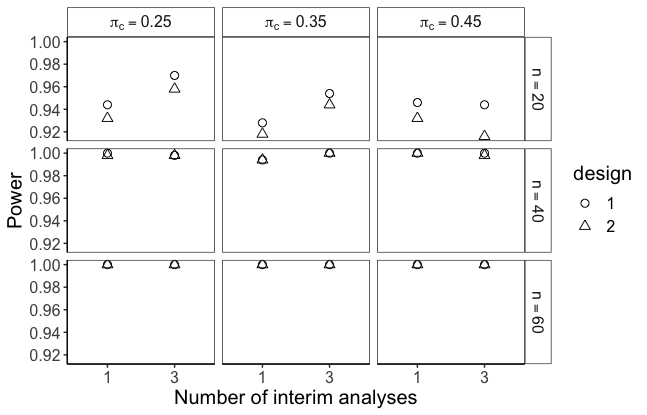}
         \caption{$\rho=0.05$, $U=0.95$, $\Delta=0.2$}
         \label{PWR-binary-multiple-rho0.05-m8-U0.95}
     \end{subfigure}
     %\hfill
     \begin{subfigure}[b]{0.49\textwidth}
         \centering
         \includegraphics[width=\textwidth]{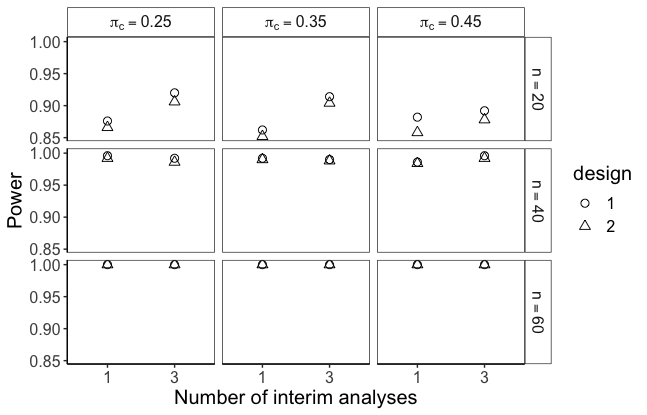}
         \caption{$\rho=0.1$, $U=0.95$, $\Delta=0.2$}
         \label{PWR-binary-multiple-rho0.1-m8-U0.95}
     \end{subfigure}\\
     \begin{subfigure}[b]{0.49\textwidth}
         \centering
         \includegraphics[width=\textwidth]{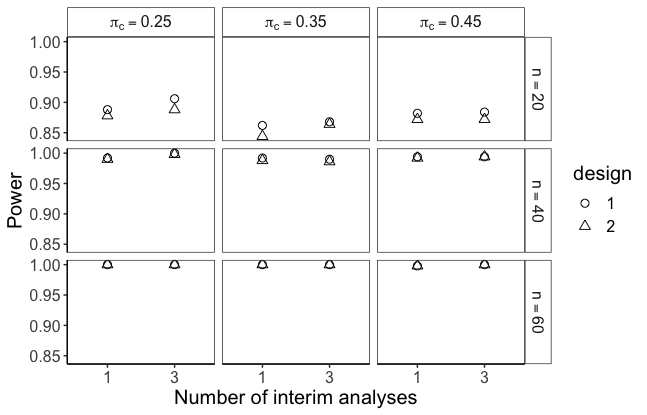}
         \caption{$\rho=0.05$, $U=0.98$, $\Delta=0.2$}
         \label{PWR-binary-multiple-rho0.05-m8-U0.98}
     \end{subfigure}
     %\hfill
     \begin{subfigure}[b]{0.49\textwidth}
         \centering
         \includegraphics[width=\textwidth]{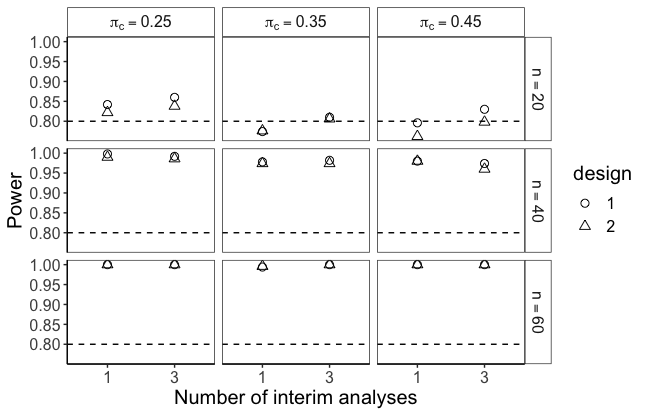}
         \caption{$\rho=0.1$, $U=0.98$, $\Delta=0.2$}
         \label{PWR-binary-multiple-rho0.1-m8-U0.98}
     \end{subfigure}\\
        \caption{Plot of power versus number of interim looks for $n=20, 40, 60$, $\pi_c = 0.25, 0.35, 0.45$, $m=8$ with subpanels (a)-(h) indicating all possible combinations of $\rho = 0.05, 0.1$, $U=0.95, 0.98$ and $\Delta=0.1, 0.2$. The dashed lines show the power of 0.8.}
        \label{PWR-binary-multiple}
\end{figure}

\begin{figure}[H]
     \centering
     \begin{subfigure}[b]{0.49\textwidth}
         \centering
         \includegraphics[width=\textwidth]{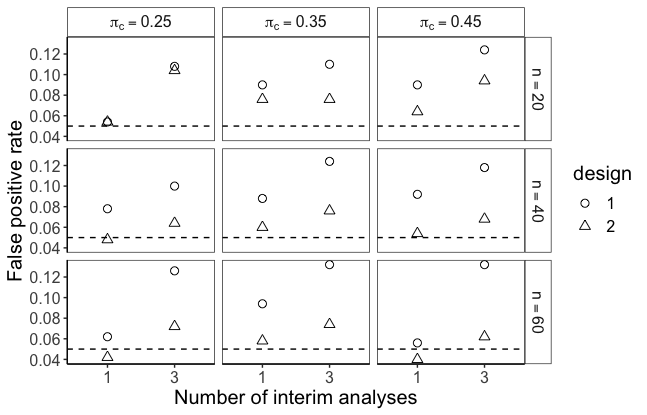}
        \caption{$\rho=0.05$, $U=0.95$}
        \label{FPR-binary-multiple-rho0.05-m16-U0.95}
     \end{subfigure}
     %\hfill
     \begin{subfigure}[b]{0.49\textwidth}
         \centering
         \includegraphics[width=\textwidth]{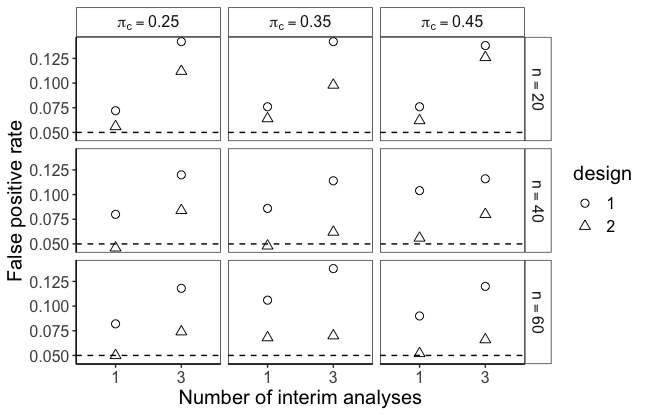}
         \caption{$\rho=0.1$, $U=0.95$}
         \label{FPR-binary-multiple-rho0.1-m16-U0.95}
     \end{subfigure}\\
     \begin{subfigure}[b]{0.49\textwidth}
         \centering
         \includegraphics[width=\textwidth]{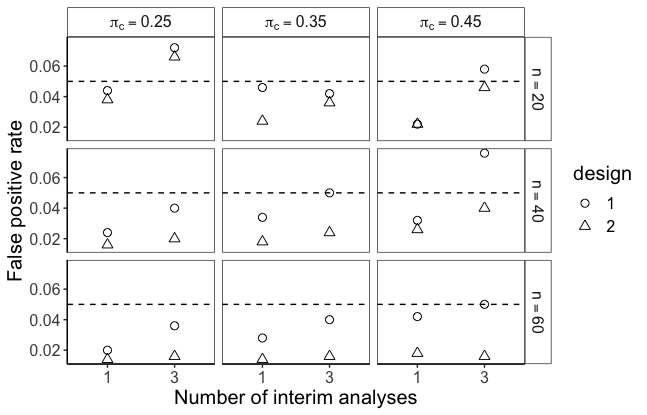}
        \caption{$\rho=0.05$, $U=0.98$}
         \label{FPR-binary-multiple-rho0.05-m16-U0.98}
     \end{subfigure}
     %\hfill
     \begin{subfigure}[b]{0.49\textwidth}
         \centering
        \includegraphics[width=\textwidth]{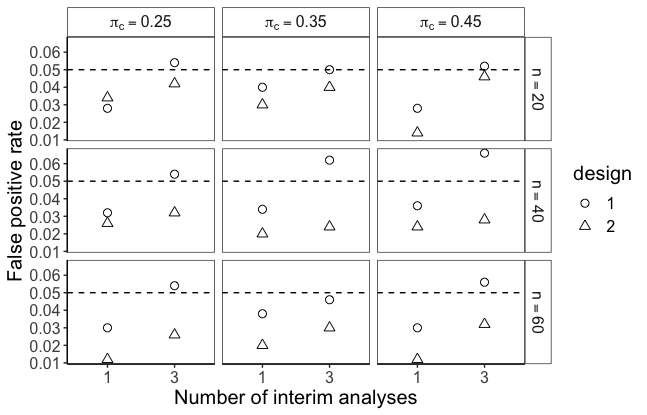}
         \caption{$\rho=0.1$, $U=0.98$}
         \label{FPR-binary-multiple-rho0.1-m16-U0.98}
     \end{subfigure}\\
        \caption{Plot of false positive rate versus number of interim looks for $n=20, 40, 60$, $\pi_c = 0.25, 0.35, 0.45$, $m=16$ with subpanels (a)-(d) indicating all possible combinations of $\rho = 0.05, 0.1$ and $U=0.95, 0.98$. The dashed lines show the false positive rate of 0.05.}
        \label{FPR-binary-multiple-m16}
\end{figure}

\begin{figure}[H]
     \centering
     \begin{subfigure}[b]{0.49\textwidth}
         \centering
         \includegraphics[width=\textwidth]{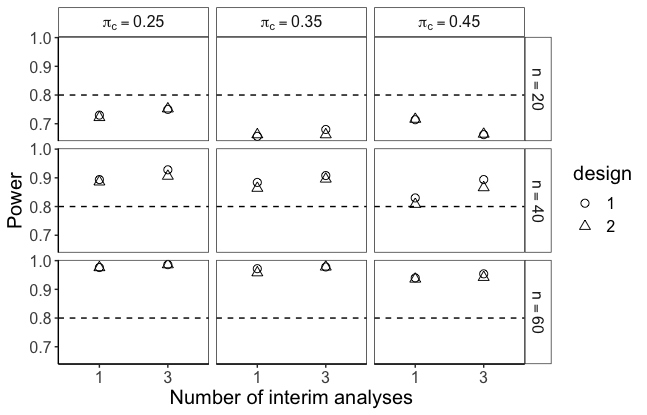}
         \caption{$\rho=0.05$, $U=0.95$, $\Delta=0.1$}
         \label{PWR-binary-multiple-delta0.1-rho0.05-m16-U0.95}
     \end{subfigure}
     %\hfill
     \begin{subfigure}[b]{0.49\textwidth}
         \centering
         \includegraphics[width=\textwidth]{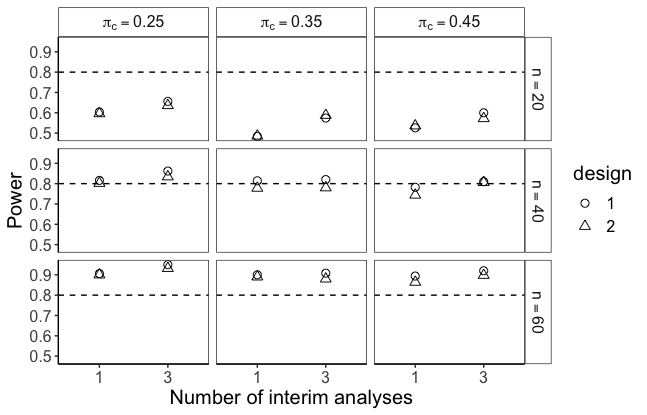}
         \caption{$\rho=0.1$, $U=0.95$, $\Delta=0.1$}
         \label{PWR-binary-multiple-delta0.1-rho0.1-m16-U0.95}
     \end{subfigure}\\
     \begin{subfigure}[b]{0.49\textwidth}
         \centering
         \includegraphics[width=\textwidth]{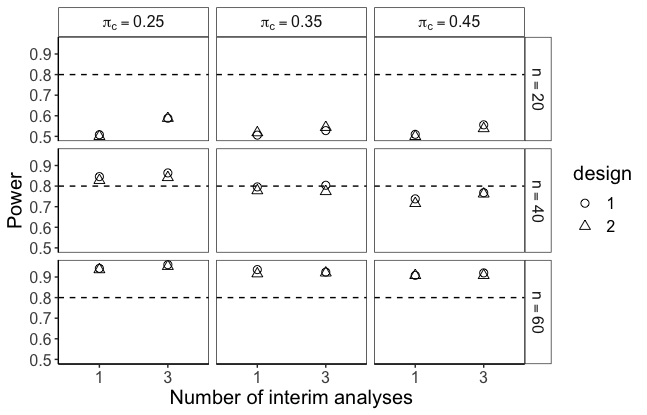}
         \caption{$\rho=0.05$, $U=0.98$, $\Delta=0.1$}
         \label{PWR-binary-multiple-delta0.1-rho0.05-m16-U0.98}
     \end{subfigure}
     %\hfill
     \begin{subfigure}[b]{0.49\textwidth}
         \centering
         \includegraphics[width=\textwidth]{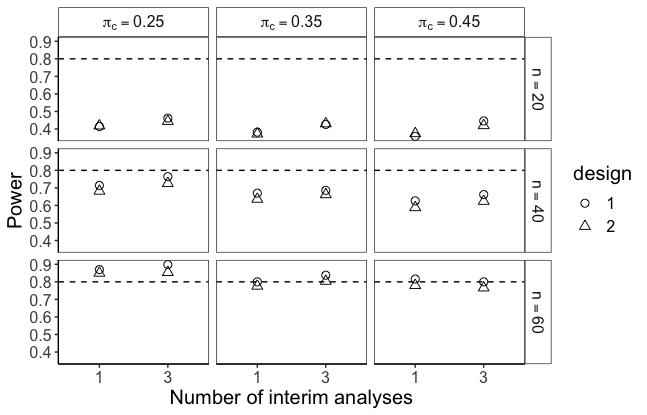}
         \caption{$\rho=0.1$, $U=0.98$, $\Delta=0.1$}
         \label{PWR-binary-multiple-delta0.1-rho0.1-m16-U0.98}
     \end{subfigure}
      \begin{subfigure}[b]{0.49\textwidth}
         \centering
         \includegraphics[width=\textwidth]{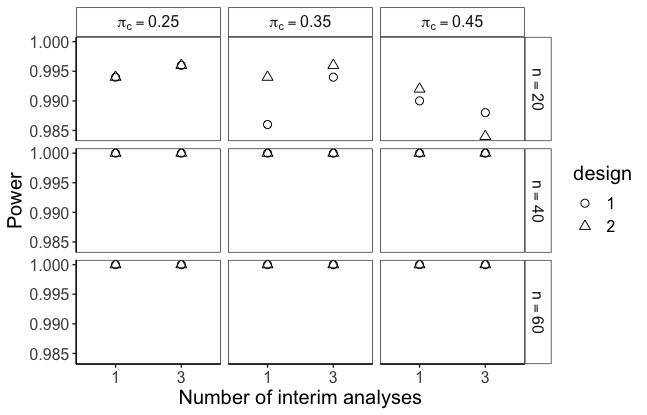}
         \caption{$\rho=0.05$, $U=0.95$, $\Delta=0.2$}
         \label{PWR-binary-multiple-rho0.05-m16-U0.95}
     \end{subfigure}
     %\hfill
     \begin{subfigure}[b]{0.49\textwidth}
         \centering
         \includegraphics[width=\textwidth]{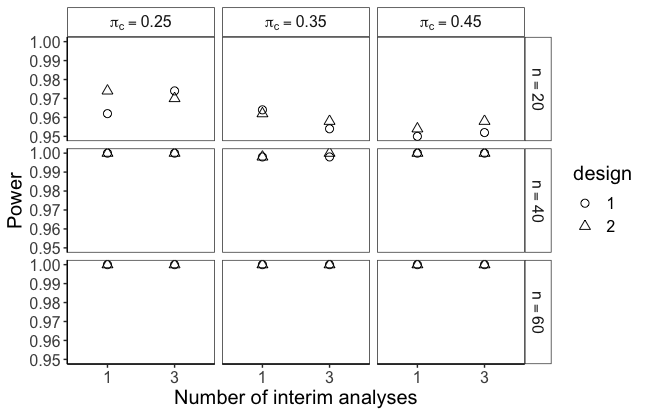}
         \caption{$\rho=0.1$, $U=0.95$, $\Delta=0.2$}
         \label{PWR-binary-multiple-rho0.1-m16-U0.95}
     \end{subfigure}
     \begin{subfigure}[b]{0.49\textwidth}
         \centering
         \includegraphics[width=\textwidth]{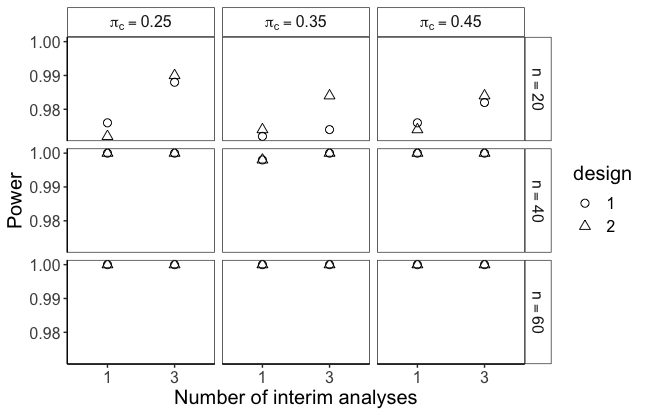}
         \caption{$\rho=0.05$, $U=0.98$, $\Delta=0.2$}
         \label{PWR-binary-multiple-rho0.05-m16-U0.98}
     \end{subfigure}
     %\hfill
     \begin{subfigure}[b]{0.49\textwidth}
         \centering
         \includegraphics[width=\textwidth]{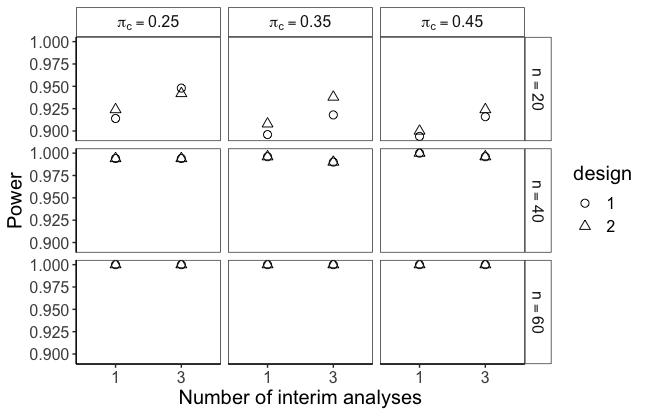}
         \caption{$\rho=0.1$, $U=0.98$, $\Delta=0.2$}
         \label{PWR-binary-multiple-rho0.1-m16-U0.98}
     \end{subfigure}
        \caption{Plot of power versus number of interim looks for $n=20, 40, 60$, $\pi_c = 0.25, 0.35, 0.45$, $m=16$ with subpanels (a)-(h) indicating all possible combinations of $\rho = 0.05, 0.1$, $U=0.95, 0.98$ and $\Delta=0.1, 0.2$. The dashed lines show the power of 0.8.}
        \label{PWR-binary-multiple-m16}
\end{figure}

The design performance for multiple interim analyses is shown in Figures \ref{FPR-binary-multiple} and \ref{PWR-binary-multiple}. %For power, only results for an effect size of 0.1 is presented, and the results for an effect size of 0.2 is displayed in Figure \ref{PWR-binary-multiple-delta0.2} in Appendix \ref{secA2}.
Design 1 still has higher false positive rate and the powers for the two designs are similar over the parameter space we explored. With multiple interim analyses, false positive rate will increase for most cases.  The increase in power, when $n=20$ and $\Delta=0.1$, outweighs the increase in false positive rate. However, in this setting the powers for both designs are still unsatisfactory with each possible combination of ICC, decision boundary or baseline risk we explored. It suggests that when the effect size is 0.1, 20 clusters for each treatment arm may not be sufficient to ensure the desired power. When we have more clusters, such as 40 or 60, with multiple interim analyses, the increase in power is relatively slight compared with the increase in false positive rate for most cases. When the effect size is 0.2, a smaller sample size is sufficient to ensure the desired power. Therefore, the power is satisfactory for almost all scenarios we investigated.
%For design 1, the increase in power is relatively slight compared with the increase in false positive rate while for design 2, the increases in the two measures are both quite small.
Similarly, with multiple interim analyses a larger boundary value can evidently help reduce false positive rate, especially for design 1. The resulting decrease in power is acceptable for a sufficient sample size. However, with a small sample size, a larger decision boundary may lead to insufficient power.  
 The results for a larger cluster size are also displayed in Figures \ref{FPR-binary-multiple-m16} and \ref{PWR-binary-multiple-m16}. %in Appendix \ref{subsecA32}.

Therefore, for binary outcomes, for the parameters explored in the simulation, design 2 is recommended since it has smaller false positive rate and performs comparably with design 1 in terms of power. Additionally, a single interim analysis is sufficient to control false positive rate while maintaining satisfactory power. Regarding the decision boundary, if design 2 with single interim analysis is planned, $U=0.95$ may be conservative enough for obtaining a satisfactory false positive rate as well as power. %, and in this case, a decision boundary of 0.98 will make the test over-conservative. 
However, if due to feasibility, design 1 is preferred or multiple interim analyses are of interest %needed due to the slow enrollment of participants, 
a larger decision boundary such as $U=0.98$ may be recommended.  %Also, it is recommended to use $U=0.98$ instead of $U=0.95$ as the decision boundary in stopping rule and single interim analysis is enough to keep false positive rate at relatively low while the power will not be reduced a lot. 
%However, obtaining 
Developing a more general set of recommendations requires further exploration.

\end{appendices}



\end{document}